\documentclass[prd,aps,preprint,nofootinbib,superscriptaddress]{revtex4}
\usepackage{epsfig}
\usepackage{amsmath}
\input{epsf}

\begin{document}

\preprint{BNL-NT-06/30} \preprint{RBRC-611}

\vspace*{2cm}
\title{Single transverse-spin asymmetry in high transverse momentum
pion production in $pp$ collisions \\[1cm]}

\author{Chris Kouvaris}
\email{kouvaris@nbi.dk}
\affiliation{The Niels Bohr Institute,
             Blegdamsvej 17, DK-2100 Copenhagen 0, Denmark}
\author{Jian-Wei Qiu}
\email{jwq@iastate.edu}
\affiliation{Department of Physics and Astronomy,
             Iowa State University, Ames, IA 50011}
\affiliation{Physics Department, Brookhaven National Laboratory,
             Upton, NY 11973}
\author{Werner Vogelsang}
\email{vogelsan@quark.phy.bnl.gov}
\affiliation{Physics Department, Brookhaven National Laboratory,
             Upton, NY 11973}
\author{Feng Yuan}
\email{fyuan@quark.phy.bnl.gov}
\affiliation{RIKEN BNL Research Center, Building 510A,
             Brookhaven National Laboratory, Upton, NY 11973}
\date{\today\\[1cm]}

\begin{abstract}
We study the single-spin (left-right) asymmetry in
single-inclusive pion production in hadronic scattering.
This asymmetry is power-suppressed in the transverse momentum
of the produced pion and can be analyzed in terms of twist-three
parton correlation functions in the proton. We present new
calculations of the corresponding partonic hard-scattering
functions that include the so-called ``non-derivative'' contributions
not previously considered in the literature. We find a remarkably
simple structure of the results. We also present a brief phenomenological
study of the spin asymmetry, taking into account data from
fixed-target scattering and also the latest information available
from RHIC. We make additional predictions that may be tested experimentally
at RHIC. 
\end{abstract}

\maketitle

\newcommand{\nn}{\nonumber}
\newcommand{\be}{\begin{equation}}
\newcommand{\ee}{\end{equation}}
\newcommand{\ben}{\[}
\newcommand{\een}{\]}
\newcommand{\beqn}{\begin{eqnarray}}
\newcommand{\eeqn}{\end{eqnarray}}
\def\beq{\begin{equation}}
\def\eeq{\end{equation}}
\def\beeq{\begin{eqnarray}}
\def\eeeq{\end{eqnarray}}
\newcommand{\ba}{\begin{array}}
\newcommand{\ea}{\end{array}}
\newcommand{\Tr}{{\rm Tr} }
\newcommand{\dsp}{\displaystyle}
\newcommand\Eqn[1]{Eq.~(\ref{#1})}  % includes ``Eq.'' in front
\newcommand\eqn[1]{(\ref{#1})}      % parentheses around the LaTex "ref" macro

\section{Introduction}

The single-transverse spin asymmetry in the process $pp\to
\pi X$ is among the simplest spin observables in hadronic scattering.
One scatters a beam of transversely polarized protons off unpolarized
protons and measures the numbers of pions produced to either the
left or the right
of the plane spanned by the momentum and spin directions of the initial
polarized protons.  This defines a ``left-right'' asymmetry. Equivalently,
the asymmetry may be obtained by flipping the spins of the initial
polarized protons. This gives rise to the customary definition
\beq
A_N(\ell,\vec{s}_T)\equiv \frac{\sigma(\ell,\vec{s}_T)-\sigma(\ell,-\vec{s}_T)}
{\sigma(\ell,\vec{s}_T)+\sigma(\ell,-\vec{s}_T)}\equiv
\frac{\Delta\sigma(\ell,\vec{s}_T)}{\sigma(\ell)} \; ,
\eeq
where $\vec{s}_T$ denotes the transverse spin vector and $\ell$ the
four-momentum
of the produced pion.  We have written in short the symbol $\sigma$ for the
cross section; we will be interested here in the invariant differential
cross section $E d^3 \sigma(\ell,\vec{s}_T)/d^3\ell$, where $E$ is the
produced pion's energy. We assume the pions to be
produced at large transverse momentum $\ell_\perp$. Needless to say, one can
consider analogous single-spin asymmetries with other hadrons in the initial
or final states, or with a final-state photon or hadronic jet.

Measurements of single-spin asymmetries in hadronic scattering
experiments over the past three decades have shown spectacular
results. Large asymmetries of up to several tens of per cents were
observed at forward (with respect to the polarized initial beam)
angles of the produced pion. Until a few years ago, all these
experiments were done with a polarized beam impeding on a fixed
target (see, for example~\cite{E704}). These experiments necessarily had a
relatively limited kinematic reach, in particular in $\ell_\perp$.
Now, after the advent of the first polarized-proton collider, the
Relativistic Heavy Ion Collider RHIC, it has become possible to
investigate $A_N$ at higher energies \cite{star,phenix,brahms},
in a kinematic regime where the theoretical description is
bound to be under better control.

Despite the conceptual simplicity of $A_N$, the theoretical
analysis of single-spin asymmetries in hadronic scattering is
remarkably complex. The reason for this is that the asymmetry for
a single-inclusive reaction like $p^\uparrow p\to  \pi X$ (the
symbol $\uparrow$ denoting from now on the polarization of the proton) is
power-suppressed as $1/\ell_\perp$ in the hard scale set by the
observed large pion transverse momentum. This is in contrast to
typical double (longitudinal or transverse) spin asymmetries that
usually scale for large $\ell_\perp$. In essence, the
leading-twist part cancels in the difference
$\sigma(\ell,\vec{s}_T)-\sigma(\ell,-\vec{s}_T)$ in the numerator
of $A_N$. That $A_N$ must be power-suppressed is easy to see: the
only leading-power distribution function in the proton associated
with transverse polarization is {\em transversity}
\cite{transversity}. For transversity to contribute, the
corresponding partonic hard-scattering functions need to involve a
transversely polarized quark scattering off an unpolarized one.
Cross sections for such reactions vanish in perturbative QCD for
massless quarks because they require a helicity-flip for the
polarized quark, which the perturbative $q\bar{q}g$ vertex does
not allow. In addition, a non-vanishing single-spin asymmetry requires
the presence of a relative interaction phase between the
interfering amplitudes for the different helicities. At
leading twist this phase can only arise through a loop correction,
which is of higher-order in the strong coupling constant and hence
leads to a further suppression. These arguments are, in fact, more than
30 years old~\cite{kpr} and led to the general expectation that single-spin
asymmetries should be very small, in striking contrast with the
experimental results.

Power-suppressed contributions to hard-scattering processes are
generally much harder to describe in QCD than leading-twist ones.
In the case of the single-spin asymmetry in $pp\to \pi X$,  a
complete and consistent framework could be developed,
however~\cite{qs}. It is based on a collinear factorization
theorem at non-leading twist that relates the single-spin cross
section to convolutions of twist-three quark-gluon correlation
functions for the polarized proton with the usual parton
distributions for the unpolarized proton and the pion fragmentation
functions, and with hard-scattering functions calculated
from an interference of two partonic scattering amplitudes:
one with a two-parton initial state and the other with a three-parton
initial state~\cite{qs,Efremov}.
As we shall review below, the necessary phases naturally arise
in these hard-scattering functions from the interference of the two
amplitudes~\cite{qs,Efremov}.
Other, related, contributions to the single-spin asymmetry have
been proposed as well, for which the twist-three function
is associated with the unpolarized proton, or with the fragmentation
functions \cite{Koike}.

We note that also other frameworks have been considered in the
literature for describing single-spin asymmetries in hadronic
scattering. One of these introduces distribution functions that depend on
intrinsic transverse momenta of partons inside the proton
\cite{sivers}, correlated with the proton spin. Because hadronic
cross sections are steeply falling functions of $\ell_\perp$,
relatively modest intrinsic transverse momenta may generate
substantial single-spin effects. Calculations based on
this approach have had considerable phenomenological success
\cite{Ans94}; however, they rely on a factorization in terms of
transverse-momentum-dependent (TMD) distributions that has
generally not been established so far. They should therefore perhaps
be regarded as models for the power-suppressed $A_N$, in contrast to
the framework developed in~\cite{qs} which is derived from QCD
perturbation theory\footnote{We emphasize, however, that the
situation is different in cases where a hard scale is present and
a small transverse momentum is measured. Here the  TMD
distributions are indeed important ingredients to the theoretical
description. For recent work on the connection between the
twist-three and the TMD approaches for this case,
see~\cite{jqvy}.}.

In the present paper, we extend the work of Ref.~\cite{qs}. Only a
certain class of contributions, the so-called ``derivative''
pieces, to be introduced in detail below, were considered
in~\cite{qs}. These indeed dominate in the kinematic regime of
interest for single-spin asymmetries, in particular at forward angles
of the produced pion and at the lower fixed-target energies. Here
we also derive the ``non-derivative'' contributions. The full
structure of the theoretical twist-three expression for a
single-spin asymmetry contains both the derivative and the
non-derivative contributions, and it is an interesting theoretical
question how the two contributions combine in the final result.
Furthermore, from a phenomenological point of view, the
non-derivative contributions are expected to become relevant at
more central pion production angles and also at higher energies.

All in all, our study is motivated to a large extent by the advent
of data from the RHIC collider~\cite{star,phenix,brahms}. By
establishing that large asymmetries at forward angles persist to
high energies, measurements at RHIC have already opened a new
chapter on single-spin asymmetries in hadronic scattering. We
emphasize that at RHIC also the unpolarized pion production cross
section has been measured, in the same kinematic regimes as
covered by the measurements of the single-spin
asymmetries~\cite{star,phenix,brahms,cross_star,cross_phenix,cross_brahms}.
An overall very good agreement between the data and
next-to-leading order (NLO) perturbative calculations based on
collinear factorization was found
\cite{cross_star,cross_phenix,cross_brahms}. This is in contrast
to the situation in the fixed-target regime, where a serious
shortfall of NLO theory was observed \cite{resum,resum1}. Thus, it
appears that the single-spin asymmetry data from RHIC, for the
first time, can be adequately described by theoretical
calculations based on collinear factorization and partonic
hard-scattering functions calculated to low orders in perturbation
theory. Even though the calculations described in this work are
all only at the leading-order (LO) level, we are confident that
they offer relatively solid predictions for spin asymmetries. The
prospects of more data to come in the near-term future clearly
warrant renewed and detailed theoretical calculations and studies.
We regard our paper as a significant step in that direction.

This paper is organized as follows: in Sec.~\ref{calc}, we present
our calculation of the single-spin asymmetry in hadronic
scattering. We keep the presentation as compact as possible and
refer to Ref.~\cite{qs} for some further details. We introduce the
kinematics, discuss the factorization and then present in some
detail the calculation of the partonic twist-three hard-scattering
functions. Here we focus on the new aspect of our work, the
derivation of the non-derivative contributions, for which we
find a remarkably simple structure. In
Sec.~\ref{pheno} we present a phenomenological study using our new
results. We in particular fit the unknown twist-three quark-gluon
correlation functions to the new RHIC data and to some of the
older fixed-target data. We use the fit results to make further predictions
for spin asymmetries measurable at RHIC. Finally, we draw our conclusions 
in Sec.~\ref{concl}.

\section{Calculation of single-spin asymmetry \label{calc}}

We start by specifying our notation for the kinematics. We consider
the reaction
\beq
A(P,\vec{s}_T)+B(P')\rightarrow h(\ell) +X \; ,
\eeq
where $A$ is a transversely polarized spin-1/2 hadron with momentum $P$ and
spin vector $\vec{s}_T$, $B$ is an unpolarized hadron with momentum $P'$,
and $h$ is a hadron produced with momentum $\ell$. The reaction is completely
inclusive otherwise. We define the Mandelstam variables
\beeq
\label{mandel}
S&=&(P+P')^2 \simeq 2P\cdot P'  \; , \nn \\
T&=&(P-\ell)^2 \simeq -2P \cdot \ell \; , \nn \\
U&=&(P'-\ell)^2 \simeq -2P' \cdot \ell \; ,
\eeeq
and the Feynman-variable
\beq
x_F=\frac{2\ell_z}{\sqrt{S}}=\frac{T-U}{S} \; ,
\label{xfdef}
\eeq
where the last equality holds in the hadronic center-of-mass system.

\subsection{Factorization of the spin-dependent cross section}

As was shown in Ref.~\cite{qs}, to leading power in the transverse
momentum $\ell_\perp$ of the produced hadron, the spin-dependent cross section
$d\Delta\sigma(\ell_\perp,\vec{s}_T)$ factorizes into combinations
of three-field twist-3 matrix elements, twist-2 parton distributions
and/or fragmentation functions, and partonic hard-scattering functions.
The general structure of the cross section is
\begin{eqnarray}
\Delta\sigma_{A+B\rightarrow h X}(\ell_\perp,\vec{s}_T)
&=& \sum_{abc} \phi^{(3)}_{a/A}(x_1,x_2,\vec{s}_T)
             \otimes \phi_{b/B}(x')
             \otimes H_{ab\rightarrow c}(\ell_\perp,\vec{s}_T)
             \otimes D_{c\rightarrow h}(z)
\nonumber \\
&+& \sum_{abc} \delta q_{a/A}(x,\vec{s}_T)
             \otimes \phi^{(3)}_{b/B}(x_1',x_2')
             \otimes H_{ab\rightarrow c}'(\ell_\perp,\vec{s}_T)
             \otimes D_{c\rightarrow h}(z)
\label{s3e11}\nonumber  \\
&+& \sum_{abc} \delta q_{a/A}(x,\vec{s}_T)
             \otimes \phi_{b/B}(x')
             \otimes H_{ab\rightarrow c}''(\ell_\perp,\vec{s}_T)
             \otimes D^{(3)}_{c\rightarrow h}(z_1,z_2) \nonumber   \\[2mm]
&+& \mbox{higher-power corrections\, ,}
\end{eqnarray}
where the symbol $\otimes$ denotes an appropriate convolution in
partonic light-cone momentum fractions, to be specified below. Additional
arguments, such as the pion transverse momentum or the
factorization/renormalization scales, have been suppressed.
The superscripts ``$(3)$'' in Eq.~(\ref{s3e11}) indicate the
higher-twist functions. The other functions, $\phi_{b/B}(x')$, $\delta
\phi_{a/A}(x)$ and $D_{c\rightarrow h}(z)$, are the standard
twist-two unpolarized and transversity parton distributions, and the
fragmentation functions, respectively. The sums run over all parton
flavors: quarks, anti-quarks and gluons. As Eq.~(\ref{s3e11})
shows, there are in general three types of contributions
to the cross section, distinguished by the twist-3 function being
associated with either the polarized proton (first line), the
unpolarized proton (second line), or the fragmentation process
(third line). For each of these contributions, there is a
separate set of partonic hard-scattering cross sections,
denoted by $H_{ab\rightarrow c}$,  $H_{ab\rightarrow c}'$,
$H_{ab\rightarrow c}''$ in Eq.~(\ref{s3e11}).
In this paper, we will consider only the contributions of the
first type to the spin-dependent cross section. The other two are
expected to be suppressed relative to the first one, as discussed
in~\cite{qs} and verified by explicit calculation in~\cite{Koike}.

\begin{figure}
%\hspace*{-2.5cm}
\includegraphics[height=3.5cm]{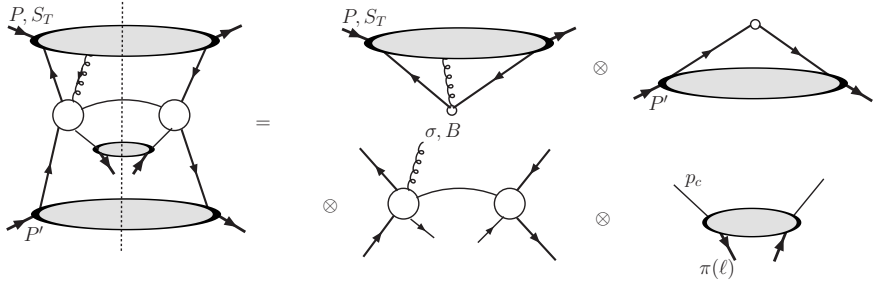}
%\vspace*{-10.5cm}
\caption{\it Generic Feynman diagram contributing to the single
transverse-spin asymmetry for inclusive pion production in
proton-proton scattering at leading twist (twist-three). The polarized
cross section can be factorized into convolutions of the
following terms: twist-three quark-gluon correlation functions
for the transversely polarized proton, parton distributions for the
unpolarized proton, pion fragmentation functions, and
hard-scattering functions calculable in QCD perturbation theory.}
\label{fig1}
\end{figure}
We can think of the contribution in the first line
of~(\ref{s3e11}) in terms of the generic Feynman diagram shown in
Fig.~\ref{fig1}. The upper part of the diagram represents a
twist-3 function for the polarized proton, generically given by a
three-parton correlation. As was discussed
in~\cite{qs}, the dominant contributions to the polarized cross
section at forward production angles of the pion are expected from
a correlation that connects two quarks and a gluon to the
hard-scattering function. We will focus on this particular
contribution as well. Other contributions, involving three
exchanged gluons \cite{gluon}, will also exist and play a possibly
important role in production at mid-rapidity.

\begin{figure}
%\hspace*{-2.5cm}
\includegraphics[height=4.0cm]{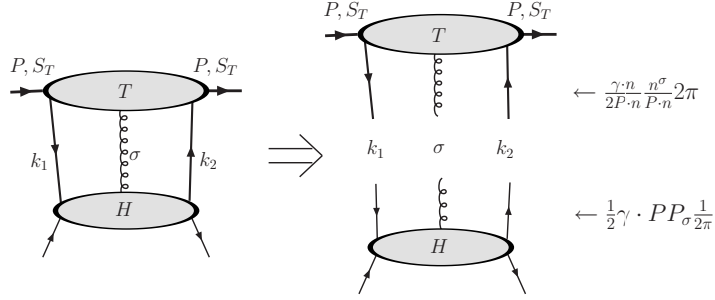}
%\vspace*{-10.5cm}
%\epsfysize=2.2in
\caption{\it Factorization of hard part and twist-three
matrix element: (a) before and (b) after separation of spinor trace and
Lorentz indices. For simplicity we have omitted the unpolarized parton
distribution and the fragmentation function. $n\sim P'$ is a light-like vector
with spatial components in direction opposite to those of the initial
momentum $P$.}
\label{fig2}
\end{figure}

In order to find the field-theoretic expression for the twist-3
function in the first line of Eq.~(\ref{s3e11}), and to
derive the rules for computing the associated hard-scattering
functions, we consider the diagram in Fig.~\ref{fig2}. Here
the parts labeled $T_a$ and $H_{ab\to c}$ represent
the twist-3 function and the partonic hard-scattering, respectively,
which are connected by the two independent
integrals over the momenta $k_1$ and
$k_2$ that they share. We thus have the following expression for
the contribution of the diagram to the spin-dependent cross
section:
\begin{equation}
d\Delta\sigma(\ell_\perp,\vec{s}_T)
\equiv \frac{1}{2S}  \sum_{abc}
  \int \frac{d^4k_1}{(2\pi)^4}\, \frac{d^4k_2}{(2\pi)^4}\,
   T_a(k_1,k_2,\vec{s}_T)\,
         H_{ab\to c}(k_1,k_2,\ell_\perp)\otimes \phi_{b/B}(x')
\otimes D_{c\rightarrow h}(z)\ ,
\label{s3e12}
\end{equation}
where $1/2S$ is a flux factor and the sum again runs over flavors. In the
above expression, spinor, color and Lorentz indices connecting the hard and
long-distance parts have already been separated, using the techniques
developed in~\cite{qs}, as sketched in Fig.~\ref{fig2}.
In a covariant gauge, the function $H_{ab\to c}(k_1,k_2)$
is contracted with
$\left(\frac{2}{N_C^2-1}\right) (t^B)_{ij}[(1/2)
\slash{\!\!\!\!P} P_{\sigma}]/(2\pi)$,
where the factor $(2\pi)$ is due
to the normalization of the twist-3 matrix element $T_a$,
$N_C=3$ is the number of colors, $B$ and $i,j$ are the color indices
of the initial gluon and quarks, respectively, and
the matrices $(t^B)_{ij}$ are the SU(3) generators
in the fundamental representation.

The next step is to perform a ``collinear'' expansion of the expression for
the diagram~\cite{QS:DYQ}.  Due to perturbative pinch singularities
of the partonic scattering diagrams~\cite{qs}, the integration in
Eq.~(\ref{s3e12}) is dominated by the phase space
where $k_i^2\sim 0$, and
we can approximate the parton momenta $k_i$ entering the hard scattering
to be on-shell and nearly parallel to the momentum $P$ of the
initial polarized proton:
\beq
\label{vexp}
k_i^{\mu}=x_i P^{\mu}+k^{\mu}_{i,\perp}+k_{i,T}^2/(x_i S)P'^{\mu}\; ,
\eeq
where the $k^{\mu}_{i,\perp}$ are perpendicular to both $P$ and
$P'$, and where $k_{i,T}^2\equiv -(k_{i,\perp}^{\mu})^2$.
The last term $\propto k_{i,\perp}^2$ in~(\ref{vexp}) can be neglected
since it is beyond the order in $k_{i,\perp}$ that we consider.
The collinear expansion enables us to reduce the four-dimensional integrals in
Eq.\ (\ref{s3e12}) to convolutions in the light-cone momentum
fractions of the initial partons. Expanding
$H_{ab\to c}$ in the partonic momenta, $k_1$ and $k_2$,
around $k_1=x_1P$ and $k_2=x_2P$, respectively, we have
\begin{eqnarray}
H_{ab\to c}(k_1,k_2) = H_{ab\to c}(x_1,x_2)
&+& \frac{\partial H_{ab\to c}}{\partial k_{1}^{\rho}}(x_1,x_2)
          \left(k_1-x_1P\right)^{\rho} \nonumber \\
&+& \frac{\partial H_{ab\to c}}{\partial k_{2}^{\rho}}(x_1,x_2)
          \left(k_2-x_2P\right)^{\rho}
 + \ldots \ .
\label{s3e13}
\end{eqnarray}
Because of~(\ref{vexp}), the derivatives in the latter equation
are in the transverse vectors $k^{\mu}_{i,\perp}$ only.
The expansion~(\ref{s3e13}), substituted into Eq.\ (\ref{s3e12}),
allows us to integrate over three of the four
components of each of the loop momenta $k_i$.
The top part of the diagram $T_a$ then becomes
a twist-three light cone matrix element, given by
\begin{eqnarray}
T_{a,F}(x_1,x_2) &=&
\int \frac{dy_1^- dy_2^-}{4\pi}
     \mbox{e}^{ix_1P^+y_1^- + i(x_2-x_1)P^+ y_2^-}
\nonumber \\
&& \quad \times
\langle P,\vec{s}_T |\bar{\psi}_a(0)\gamma^+
 \left[\epsilon^{s_T\sigma n \bar{n}}\, F_{\sigma}^{\ +}(y_2^-)\right]
 \psi_a(y_1^-) |P,\vec{s}_T\rangle \ ,
\label{s3e18}
\end{eqnarray}
where we have introduced the subscript ``$F$'' to indicate that
the matrix element involves the gluon field strength tensor
$F_{\sigma}^{\ +}$. The additional ordered exponentials of the
gauge field that make this matrix element gauge invariant have
been suppressed \cite{QS:DYQ}; Eq.~(\ref{s3e18}) as
it stands is valid in the light-cone gauge $A^+=0$.
$T_{a,F}$ is a symmetric function of its
arguments, $T_{a,F}(x_1,x_2)=T_{a,F}(x_2,x_1)$.

\subsection{Poles in hard-scattering functions and contributions to
$k_\perp$-expansion}

\begin{figure}
%\hspace*{-0.5cm}
\includegraphics[height=11cm]{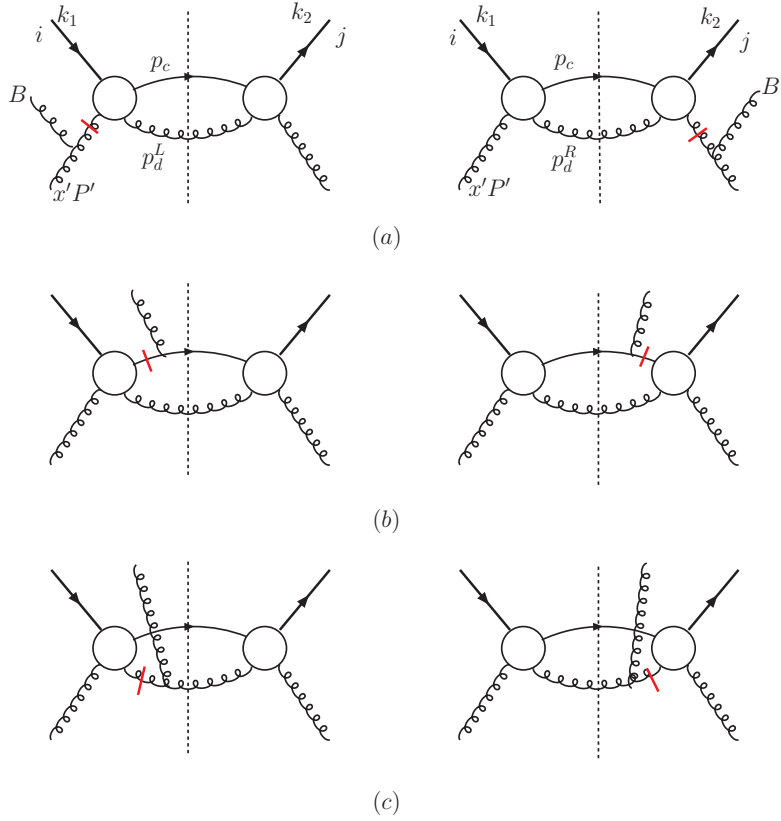}
%\vspace*{-4.5cm}
%\epsfysize=4.2in
\caption{\it Three classes of quark-gluon scattering
diagrams contributing to the spin-dependent cross section
$\Delta\sigma(\protect\vec{s}_T)$: (a) diagrams with an
initial-state pole, (b) and (c) diagrams with a final-state pole.
Symbols $B$ and $ij$ are color indices for the gluon and the
quarks. The propagator that provides the pole is indicated by a bar.
All poles shown are ``soft-gluon'' poles, contributing at $x_1=x_2$
(see text).}
\label{fig3}
\end{figure}

\begin{figure}[t]
%\hspace*{-0.5cm}
\includegraphics[height=10cm]{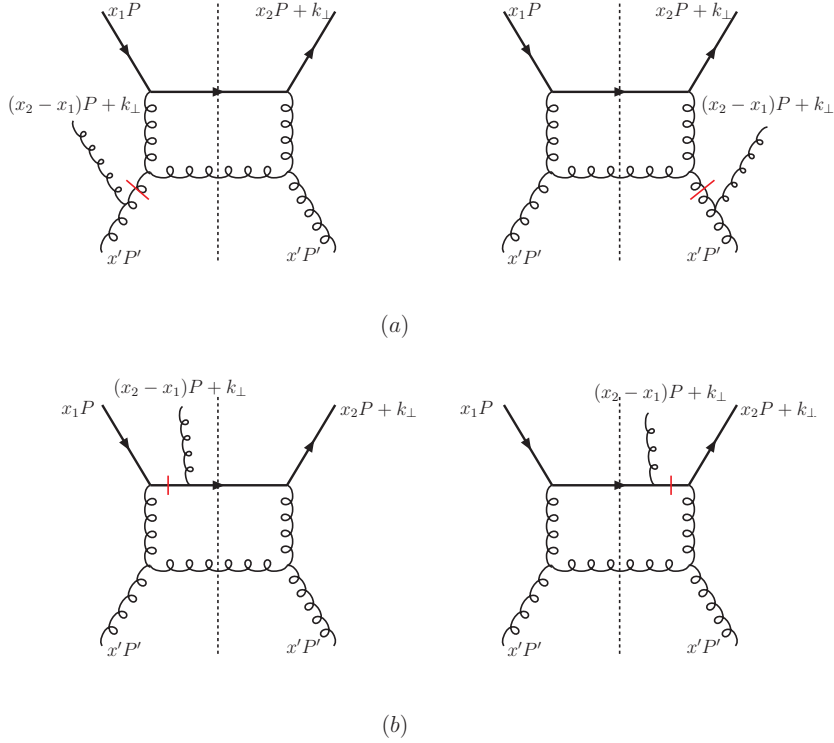}
%\vspace*{-5cm}
%\epsfysize=4.2in
\caption{\it Specific
examples of diagrams for generic (a) initial-state
and (b) final-state interactions, along with simplified notation of
external momenta.}  \label{fig4}
\end{figure}

In addition, $T_{a,F}$ is real, implying that the phase needed to
generate a single-spin asymmetry has to arise in the functions
$H_{ab\to c}$ in Eq.\ (\ref{s3e13}). As was shown in~\cite{qs,Efremov},
imaginary parts in $H_{ab\to c}$ can arise even at tree level,
thanks to the pole structure of the hard scattering function.
Examples of this are shown in Fig.~\ref{fig3} for the case
of quark-gluon scattering. Imaginary parts arise
from the scattering amplitude with an extra initial-state gluon
when its momentum integral is evaluated by the residues of
unpinched poles of the propagators indicated by the bars.
The on-shell condition associated with any such pole fixes the 
momentum fraction of the extra initial-state gluon and hence further
simplifies the integrations over the momenta $k_1$ and $k_2$ in
Eq.~(\ref{s3e12}). Roughly speaking, all of the diagrams in
Fig.~\ref{fig3} provide an unpinched pole at $x_1=x_2$,
with subtleties that we will address shortly.
At these poles, one has~\cite{qs}
\begin{equation}
\frac{\partial H_{ab\to c}}{\partial k^{\rho}_{2,\perp}}(x_1,x_2=x_1)
= - \frac{\partial H_{ab\to c}}{\partial k^{\rho}_{1,\perp}}(x_1,x_2=x_1)\, .
\label{s3e14}
\end{equation}
Thanks to this property, one can organize the calculation of the
partonic hard-scattering functions with a simpler momentum
flow, using a single transverse momentum $k_\perp$,
as shown in Fig.~\ref{fig4}.

In order to demonstrate the emergence of a strong-interaction phase
through a pole contribution at $x_1=x_2$, let us consider the specific example
for the initial-state interaction shown in Fig.~\ref{fig4}(a). We need
to consider contributions for which the initial-state gluon attaches on
the right or on the left side of the cut.
The propagator denoted by a bar in the left part of the figure reads
\beeq
\frac{1}{(x'P'+(x_2-x_1)P+k_\perp)^2+i \epsilon}
&=&\frac{1}{(x_2-x_1)x'S+i \epsilon} +{\cal O}(k_T^2)\nonumber \\
&\rightarrow& -\frac{i\pi}{x'S} \delta (x_2-x_1)  \; ,
\label{ISdelta}
\eeeq
where in the second line we have extracted the imaginary part provided
by the propagator, which contributes to the single-spin asymmetry.
When the gluon attaches on the right-hand side of the cut, we
obtain the same result, but with opposite sign. Therefore, effectively
the difference of the two diagrams in Fig.~\ref{fig4}(a) contributes.
The first term on the right-hand side of Eq.~(\ref{s3e13}) cancels
in this difference, so that eventually only the other two contribute.

In Eq.~(\ref{ISdelta}) we have neglected a term $\propto k_T^2$ since
we are only interested in first-order (linear) $k_T$ effects. A linear
term is not present in the delta-function
in~(\ref{ISdelta}) because the vector $k_\perp$
is perpendicular to both $P$ and $P'$. For final-state interaction,
this situation changes. Generic diagrams with final-state interactions
involving the ``observed'' parton
are shown in Fig.~\ref{fig4}(b). On the left side of the diagram, a
phase from the propagator marked by the bar arises as
\beeq
\frac{1}{(\ell/z-(x_2-x_1)P-k_\perp)^2+i \epsilon}
&=&\frac{1}{-2 P\cdot \ell\, (x_2-x_1)/z-2 \ell\cdot k_\perp/z+i \epsilon}
+{\cal O}(k_T^2)\nonumber \\
&\rightarrow& -\frac{i\pi z}{T} \delta (x_2-x_1-2 \ell\cdot k_\perp/T)  \; ,
\label{FSdelta}
\eeeq
where the momentum of the fragmenting (``observed'') final-state parton
is related to that of the produced hadron by $\ell=z p_c$,
and $T$ has been defined in Eq.~(\ref{mandel}). Again, the propagator
on the right side of the cut has the same pole, with opposite sign.
As Eq.~(\ref{FSdelta}) shows, the pole provided by the
final-state interactions is located near $x_1=x_2$, but displaced
by a term linear in $k_\perp$. When inserted into the collinear
expansion~(\ref{s3e13}), this term will make a contribution to the
single-spin asymmetry involving a derivative of the delta-function
and hence, by partial integration, a {\it derivative of the twist-three
quark-gluon correlation function}~\cite{qs}.

Such ``derivative'' terms may, however, also arise in a different way, through
the on-shell condition for the unobserved final-state parton.
For the diagrams on the left-hand-side of Figs.~\ref{fig4}(a) and (b),
the momentum carried by that parton is
$p_d^L=x'P'+x_2 P+k_\perp - \ell/z$, where the superscript
$L$ ($R$ introduced later) refers to the diagrams
whose extra initial-state gluon is attached to the left
(right) of the cut.  The phase space provides a delta-function
that puts this particle on its mass-shell,
\beeq
\label{x2df}
\delta \left( (p_d^L)^2 \right) &=& \delta (x_2 (x'S+T/z) +x'U/z -
2 \ell\cdot k_\perp/z) \nonumber \\
&=& \frac{1}{x'S+T/z}\delta\left(x_2 -x - \frac{2 \ell\cdot
k_\perp/z}{x'S+T/z} \right)\; , \eeeq where \beq x\equiv
\frac{-x'U/z}{x'S+T/z} \; . \label{xdef} \eeq $x$ can be
interpreted as the ``usual'' value of the partonic momentum
fraction of the polarized proton if there is no $k_\perp$.
Eq.~(\ref{x2df}) fixes $x_2$ in terms of the Mandelstam variables
$S,T,U$ and a linear term in $k_\perp$. The latter will give rise
to ``derivative'' contributions in the same way as
Eq.~(\ref{FSdelta}) does. If the initial gluon
attaches on the right-hand-side of the cut, however, the momentum
of the unobserved parton is $p_d^R=x'P'+x_1 P - \ell/z$, and the
resulting on-shell condition fixes $x_1$: \beq \delta \left(
(p_d^R)^2 \right) = \delta (x_1 -x) \; , \eeq with no dependence
on $k_\perp$.

Additional contributions to the collinear
expansion can of course also arise from terms linear in $k_\perp$
in the other ``hard'' propagators or in the numerator of each diagram.
These terms do not lead to ``derivative'' contributions to the
single-spin asymmetry, but to contributions involving
$T_{a,F}$ itself.

We close this section with two further observations. First, we note that
final-state interactions involving the ``unobserved'' parton $d$
cancel when summing over contributions where the additional gluon
attaches on the right or the left side of the cut. Second, the
contributions we have discussed are all characterized by the additional
initial gluon becoming soft. The poles arising from this are, therefore,
customarily referred to as ``soft-gluon poles''. The hard-scattering
diagrams will in general possess also other poles, for which an initial
{\it quark} becomes soft~\cite{qs}. Such ``soft-fermion poles'' are
expected to play a less important role and are not considered
in this work.

In the next section, we will provide a ``master formula'' that
allows to take into account all contributions to the $k_\perp$-expansion
discussed above simultaneously and in a systematic and relatively
straightforward manner.

\subsection{``Master formula''}

As was shown in~\cite{qs}, the factorized expression for
$\Delta\sigma(\vec{s}_T)$ takes the form
\begin{eqnarray}
d\Delta\sigma(\vec{s}_T) &\propto& \frac{1}{2S} \sum_{abc}
\int dz\, D_{c\rightarrow h}(z)
\int \frac{dx'}{x'}\, \phi_{b/B}(x')
\int dx_1 dx_2\, T_{a,F}(x_1,x_2)
\nonumber \\
&& \quad \times
 i\epsilon^{\rho s_T n \bar{n}}\,
 \lim_{k_\perp\to 0}\frac{\partial}{\partial k_\perp^{\rho}}
 H_{ab\rightarrow c}(x_1,x_2,x',z)
\ ,
\label{s3e21}
\end{eqnarray}
where \beq \epsilon^{\rho s_Tn \bar{n}}=\epsilon^{\rho \sigma \mu
\nu}s_{T\sigma}n_{\mu}n_{\nu} \eeq with $n$ and $\bar{n}$ two
light-like vectors whose spatial components are parallel to those
of $P'$ and $P$, respectively. According to the discussion in the
previous section, we are therefore led to consider the following
general expression: \beeq \label{master1} &&\lim_{k_\perp\to
0}\frac{\partial}{\partial k_\perp^{\rho}} \int dx_1 \int dx_2
\,T_{a,F}(x_1,x_2) \Big[\, H_L(x_1,x_2,k_\perp) \,\delta (
x_1-x_2+v_1 \cdot k_\perp)\delta (x_2-x-v_2 \cdot
k_\perp) \nonumber \\
&&\hspace*{6.1cm} - H_R(x_1,x_2,k_\perp)\,\delta (
x_1-x_2+v_1 \cdot k_\perp)\delta (x_1-x)\Big] \; .
\eeeq
Here, $H_L$ and $H_R$ denote the contributions to $H_{ab\to c}$
for any diagram, when the initial gluon attaches on the left or the right
side, respectively. $v_1$ and $v_2$ are
vectors made of $P$, $P'$, and $\ell$ whose form follows
directly from the preceding discussion. The first delta-function
in each of the two terms associated with $H_{L,R}$ results from
the propagator poles discussed in Eqs.~(\ref{ISdelta})
and~(\ref{FSdelta}). For initial-state interactions, $v_1=0$ (see
Eq.~(\ref{ISdelta})), for final-state ones,
$v_1=2 \ell/T$ (see Eq.~(\ref{FSdelta})). An important point is
that $v_1$ is the same vector on the left and on the right-hand-side
of the cut. The second delta-function in each term results from
the on-shell condition for the unobserved particle. As explained
earlier, these delta functions differ for the two sides of the cut.
We have $v_2=2 \ell/(x'zS+T)$ for the left side and $v_2=0$ for the right one,
which we have already used.

A straightforward way of dealing with the expression in~(\ref{master1})
is to use the various delta-functions to perform the integrations
over $x_1$ and $x_2$. This gives:
\beeq
&&\lim_{k_\perp\to 0}\frac{\partial}{\partial k_\perp^{\rho}}\Big[
T_{a,F}(x+(v_2-v_1) \cdot k_\perp,x+v_2\cdot k_\perp) \,
H_L(x+(v_2-v_1)\cdot  k_\perp,x+v_2\cdot k_\perp,k_\perp)
\nonumber \\
&&\hspace*{2cm} -\,T_{a,F}(x,x+v_1\cdot
k_\perp) \,H_R(x,x+v_1\cdot k_\perp,k_\perp)\Big]\; .
\eeeq
This term can be organized as
\beeq
\label{master2}
&&\hspace*{-9mm}(v_2-v_1)_\rho\,H_L(x,x,0)\,\frac{d T_{a,F}(x,x)}{d x}\,
+\,T_{a,F}(x,x) \,\times\\
&&\hspace*{-0.8cm} \times\,\lim_{k_\perp\to 0}
\frac{\partial}{\partial k_\perp^{\rho}}\,
\Big[ H_L(x+(v_2-v_1)\cdot k_\perp,x+v_2\cdot
k_\perp,k_\perp)-H_R(x,x+v_1\cdot k_\perp,k_\perp)
\Big]_{k_\perp=0} \; .\nonumber
\eeeq
This is our ``master formula''.  In deriving it
we have used that the hard-scattering functions with the gluon attaching
on the left or the right side of the cut are the same at $k_\perp=0$,
\beq
H_L(x,x,0)=H_R(x,x,0) \; . \label{LeqR}
\eeq

Equation~(\ref{master2})  applies to both initial- and final-state
interactions. As one can see, the first term is proportional to
$d T_{a,F}(x,x)/dx$ [thanks to Eq.~(\ref{LeqR}) it does
not matter whether we write $H_L(x,x,0)$ or $H_R(x,x,0)$
in this term]. This is the ``derivative'' contribution that we
discussed above and that was originally computed in Ref.~\cite{qs}.  The
second term involves only $T_{a,F}(x,x)$, without a derivative.
Equation~(\ref{master2})  allows a simultaneous computation of
both the derivative and non-derivative contributions.

The next step is to consider all contributing partonic reactions
and to calculate the contributions to Eq.~(\ref{master2}).
The partonic channels we need to consider are
$(qg)g\to qg$, $(qg)g\to gq$, $(qg)\bar{q}\to gg$, $(qg)q'\to qq'$,
$(qg)q'\to q'q$,
$(qg)q\to qq$, $(qg)\bar{q}\to q'\bar{q}'$, $(qg)\bar{q}\to q\bar{q}$,
where for each the first two initial partons are entering from the
polarized proton via the twist-three correlation function $T_{a,F}$.
We remind the reader that we are ignoring contributions involving a
three-gluon twist-three correlation function, which would correspond to a
$(gg)$ initial state. Crossed channels are implicit and taken into account
as well.

Upon calculating all associated hard-scattering functions, we
found that they possess a remarkable property: for each process,
\beeq \label{master3} &&\hspace*{-4mm}\lim_{k_\perp\to 0}
\frac{\partial}{\partial k_\perp^{\rho}}\, \Big[
H_L(x+(v_2-v_1)\cdot k_\perp,x+v_2\cdot
k_\perp,k_\perp)-H_R(x,x+v_1\cdot k_\perp,k_\perp)
\Big]_{k_\perp=0} \nonumber \\[2mm]
&&=-\frac{(v_2-v_1)_{\rho}}{x}\,H_L(x,x,0)\; , \eeeq which is
again valid for the case of both initial- and the final-state
interactions. We have not been able to develop a proof why
Eq.~(\ref{master3}) holds in general, even though the equation is
certainly not accidental and such a proof should be possible. In
any case,  Equation~(\ref{master3}) leads to a dramatic
simplification of the final result. Inserting~(\ref{master3})
into~(\ref{master2}), one finds the expression \beq
\label{master5} -\frac{(v_2-v_1)_\rho}{x}\,H_L(x,x,0)\,\left[
T_{a,F}(x,x) -x  T'_{a,F}(x,x)\right]\; , \eeq where we have used
the short-hand notation $T'_{a,F}(x,x)\equiv dT_{a,F}(x,x)/dx$.
Thus, even though there could have in principle been two separate
hard-scattering functions multiplying $T_{a,F}(x,x)$ and $T'_{a,F}(x,x)$,
the final result for the combined derivative and non-derivative
terms will have a {\it single} hard-scattering function for each
process, summed over initial- and final-state
contributions and multiplying simply the combination $T_{a,F}(x,x)
-x  T'_{a,F}(x,x)$. This hard-scattering function is furthermore
identical to the one calculated for the derivative piece in~\cite{qs}.
The emerging structure is then very akin to that of the
unpolarized cross section. We are now in the position to give the
final answer for the single-spin asymmetry.

\subsection{Final result}

For definiteness, we recall the expressions for the vectors
$v_1$ and $v_2$ introduced above. We have
\beq
v_2=\frac{2\ell}{x'zS+T}=-\frac{2p_c x}{\hat{u}} \; ,
\eeq
with the partonic Mandelstam variable $\hat{u}=(p_c-p')^2=x'U/z$.
For initial-state interactions, see Eq.~(\ref{ISdelta}),
we have $v_1=0$, while for final-state ones, see Eq.~(\ref{FSdelta}),
\beq
v_1= \frac{2\ell}{T} = \frac{2p_c x}{\hat{t}}\; , \;\;\;\mathrm{or} \;\;\;
v_2-v_1 =  -\frac{2p_c x}{\hat{u}}\left( 1 + \frac{\hat{u}}{\hat{t}}
\right) \; ,
\label{factor}
\eeq
where $\hat{t}=(p_c-p)^2=xT/z$.

Using~(\ref{s3e21}) and following the steps presented in detail
in Ref.~\cite{qs}, we then find the final expression for the polarized
cross section:
\begin{eqnarray}
\label{finalcr}
E_\ell\frac{d^3\Delta\sigma(\vec{s}_T)}{d^3\ell}
&=& \frac{\alpha_s^2}{S}\, \sum_{a,b,c}
    \int_{z_{\rm min}}^1 \frac{dz}{z^2}\, D_{c\to h}(z)
   \, \int_{x'_{\rm min}}^1 \frac{dx'}{x'}\, \frac{1}{x'S + T/z} \,
\phi_{b/B}(x') \\
&\times& \sqrt{4\pi\alpha_s}\,
    \left(\frac{\epsilon^{\ell s_T n \bar{n}}}{z\hat{u}}\right)
\,   \frac{1}{x}
\left[T_{a,F}(x,x)-x\left(\frac{d}{dx}T_{a,F}(x,x)\right)\right]
    H_{ab\rightarrow c}(\hat{s},\hat{t},\hat{u})\ ,
\nonumber
\end{eqnarray}
where $x$ has been defined in Eq.~(\ref{xdef}), and where
\beq
x'_{\rm min}= \frac{-T/z}{S+U/z} \;  , \;\;\;\;\; z_{\rm min}=
-\frac{T+U}{S} \; .
\eeq
The $H_{ab\rightarrow c}$ are the final hard-scattering functions
and read
\beq
H_{ab\rightarrow c}\,=\,
H^I_{ab\rightarrow c}(\hat{s},\hat{t},\hat{u}) +
        H^F_{ab\rightarrow c}(\hat{s},\hat{t},\hat{u}) \,\left(1+
        \frac{\hat{u}}{\hat{t}}\right) \; ,
\label{Hfin}
\eeq
where $H^I_{ab\rightarrow c}$ ($H^F_{ab\rightarrow c}$)
denote the contributions due to the initial-state (final-state)
interactions. The factor $(1+\hat{u}/\hat{t})$ results
from the expression for $v_2-v_1$ in Eq.~(\ref{factor}). We have
collected all $H^I_{ab\rightarrow c}$ and $H^F_{ab\rightarrow c}$
in Appendix~A. Thanks to the structure we have found,
they must coincide with the hard-scattering functions calculated for the
derivative part in Ref.~\cite{qs}, which they do, up to trivial
corrections we found for some of the color factors in~\cite{qs}.
The results presented in Appendix~A are also in a more
compact and transparent notation.

We emphasize again the simplicity of the structure in Eq.~(\ref{finalcr}),
which is very similar to that of the unpolarized cross section in the
denominator of the spin asymmetry. The
latter reads:
\begin{eqnarray}
\label{finalcrU}
E_\ell\frac{d^3\sigma}{d^3\ell} &=&
\frac{\alpha_s^2}{S}\, \sum_{a,b,c}
    \int_{z_{\rm min}}^1 \frac{dz}{z^2}\, D_{c\to h}(z)
   \, \int_{x'_{\rm min}}^1 \frac{dx'}{x'}\, \frac{1}{x'S + T/z} \,
\phi_{b/B}(x') \\
&\times& \,   \frac{1}{x} \phi_{a/A}(x)\,
    H^U_{ab\rightarrow c}(\hat{s},\hat{t},\hat{u})\ ,
\nonumber
\end{eqnarray}
with unpolarized hard-scattering functions $H^U_{ab\rightarrow c}$ and
the usual unpolarized parton distribution functions in hadron $A$,
$ \phi_{a/A}(x)$. We give the well-known~\cite{gro}
$H^U_{ab\rightarrow c}$ also in Appendix~A.

We finally note that we have written the hard-scattering functions
for both the spin-dependent and for the unpolarized case
as dimensionless functions. The power-suppression of the
single-spin asymmetry is then explicitly visible by the denominator
$\hat{u}$ in Eq.~(\ref{finalcr}). Furthermore, note the factor
$\sqrt{\alpha_s}$
in that equation, which results from the additional interaction with a
gluon field in the hard-scattering functions for the single-spin case.

\section{Phenomenological study \label{pheno}}

We now present some first numerical results for the single-spin asymmetry
derived in the previous section. We do not aim at a full-fledged analysis of
all the hadronic single-spin data at this point, but would like to examine a
few of the salient features of the new RHIC data and of the earlier E704
fixed-target pion production data. We reserve a more detailed analysis to a
future publication.

Let us begin by specifying the main ingredients to our
calculations. We first remind the reader that all our calculations
of the hard-scattering functions are only at the LO level. We
therefore use LO parton distribution and fragmentation functions
throughout, as well as the one-loop expression for the strong
coupling constant. For the unpolarized cross section we use the LO
CTEQ5L parton distribution functions \cite{cteq5l}. Our choice for
the fragmentation functions are the LO functions presented in
Ref.~\cite{kretzer}. These have the advantage that they provide
separate sets for positively and negatively charged pions, which
are needed for the comparison to the experimental data.

For the present study, we will make rather simple models for the
twist-three quark-gluon correlation functions $T_{a,F}(x,x)$
($a=u,\bar{u},d,\bar{d},s,\bar{s}$), relating
them to their unpolarized leading-twist counterparts. We recall that
we do not include any purely gluonic twist-three correlation functions,
even though we do take into account the gluon-gluon scattering
contribution in the unpolarized cross section in the denominator
of $A_N$. Our ansatz for the correlation functions is simply:
\begin{equation}
T_{a,F}(x,x,\mu) = N_a\, x^{\alpha_a}\, (1-x)^{\beta_a}\,
\phi_a (x,\mu) \; ,
\label{ansatz}
\end{equation}
where $\phi_a (x,\mu)$ is the usual twist-two parton distribution
for flavor of type $a$ in a proton.
Note that we have now written out the dependence of the functions on a
factorization scale $\mu$, which we will always choose as $\mu=\ell_\perp$.
We will in fact assume that the functions $T_{a,F}(x,x,\mu)$ evolve
in the same way as the corresponding
unpolarized leading-twist distributions. This will
certainly not be correct in general, because of the different twist of
the two types of distributions, but may be hoped to be a reasonable
assumption at the moderate to relatively large $x$ we are interested in here.

We determine the parameters in Eq.~(\ref{ansatz}) through
a ``global'' fit to experimental data for $A_N$ as functions
of $x_F$ defined in Eq.~(\ref{xfdef}), using the
expressions in Eqs.~(\ref{finalcr}) and~(\ref{finalcrU}). Here we choose
the fixed-target scattering data at $\sqrt{S}\approx 20$~GeV
by the E704 experiment~\cite{E704}
for $p^\uparrow p\to \pi^{0,\pm}X$ and $\bar{p}^\uparrow p\to \pi^{0,\pm}X$,
and the latest preliminary RHIC data at $\sqrt{S}=200$~GeV
by the STAR~\cite{star}
[for $p^\uparrow p\to \pi^{0,\pm}X$] and BRAHMS~\cite{brahms}
[for $p^\uparrow p\to \pi^{\pm}X$ and $p^\uparrow p\to K^{\pm}X$]
collaborations. The perturbative hard-scattering expression we have derived
in the previous section is expected to be only applicable at high transverse
momentum, starting from $\ell_\perp\gtrsim {\mathrm{a \; few\; GeV}}$.
Most of the available data points for $A_N$ are at $\ell_\perp$-values not
much greater than 1~GeV, however. In case of the RHIC data, we
always use the correct value for $\ell_\perp$ for each data point,
keeping however only points with $\ell_\perp>1$~GeV. For the E704
data the situation is more complicated as most of the data points
have $\ell_\perp \lesssim 1$~GeV. In addition, as we discussed in the
Introduction, there is generally a problem with the description
of even the unpolarized cross sections in the fixed-target regime,
when hard-scattering calculations at low orders of perturbation
theory are used. All-order resummations~\cite{resum1} may be very
relevant here, which are likely to affect the spin-dependent
and the unpolarized cross section in different ways.
In view of this, we are tempted to exclude the E704 data from
our analysis. On the other hand, the information on single-spin
asymmetries is overall still rather sparse, and any information
is potentially helpful. In particular, data for anti-proton
scattering are only provided by the E704 experiment. Therefore,
in order to include the E704 data in the fit,
we {\it choose} $\ell_\perp=1.2$~GeV for these data. In addition,
we allow a large shift of the overall normalization of 
the theory result used for the comparison to these data.
This shift is meant to represent in particular the possibly
large higher-order effects on $A_N$ just described.

We have performed two separate fits to the data. One is
a ``two-flavor'' fit, for which we use only the two valence
densities $u_v$ and $d_v$ in the ansatz~(\ref{ansatz}) and set
all other distributions to zero. For this fit we
introduce a normalization factor $N_{\mathrm{E704}}=0.5$
for the theory asymmetries in the kinematic region
of the E704 data and find:
\begin{eqnarray}
{\mathrm{Fit\; I:}\;\;\;\;}
&& N_{u_v} = 0.275  \; , \;\;\; N_{d_v} = -0.365 \; ,
 \nonumber \\
&& \alpha_{u_v} = 0.508 \; , \;\;\;  \beta_{u_v} = 0.399 \; , \;\;\;
\alpha_{d_v} = -0.108 \; , \;\;\; \beta_{d_v} = 0.287 \; .
\label{fitres1}
\end{eqnarray}
\begin{figure}[b]
\hspace*{-0.5cm} \epsfig{figure=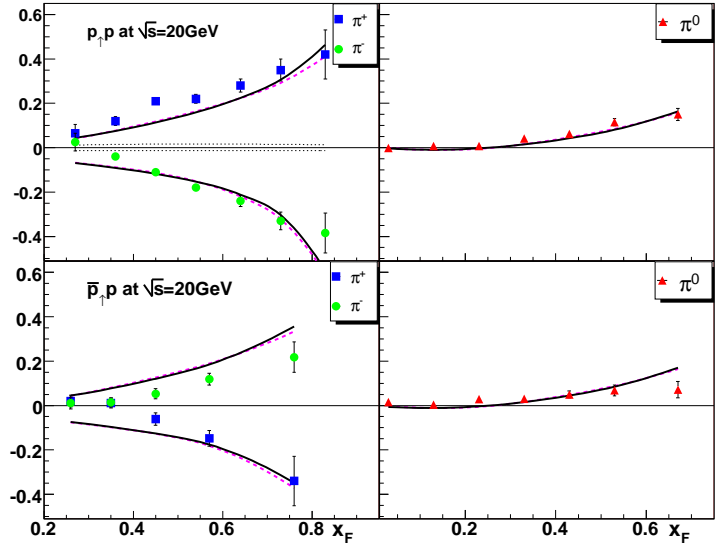,width=0.6\textwidth}
\caption{\it Comparison of the single-spin asymmetries $A_N$ using
our fit results in Eqs.~(\ref{ansatz}),(\ref{fitres1}),(\ref{fitres2}) 
to the data from E704~\cite{E704}. The solid lines are for Fit~I
(Eq.~(\ref{fitres1})), and the dashed ones are for Fit~II
(Eq.~(\ref{fitres2})). The lower dotted lines in the upper left
part of the figure show the contributions to $A_N$ for $\pi^\pm$
production by the ``non-derivative'' terms alone, for Fit~I. Note
that the theory curves in the figure are normalized by
$N_{\mathrm{E704}}=0.5$.} \label{fig5a}
\end{figure}
\noindent
The fit has a $\chi^2$-value of $304.6$ for the 60 data points and
is therefore of rather poor quality. Nonetheless, as one can see
from the comparison of the fit to the experimental data shown by
the solid lines in Fig.~\ref{fig5a} and~\ref{fig5b},
it has overall the right qualitative features.
For example, for $p^\uparrow p\to \pi^+X$ the asymmetry is positive,
which is reflected in a positive valence-$u$ twist-three correlation
function emerging from the fit. Likewise, the fact that $A_N$ is
negative for $p^\uparrow p\to \pi^- X$ implies a negative valence-$d$
distribution. For anti-proton scattering, the respective asymmetries are then
necessarily opposite, because one has~\cite{qs}
\beq
T_{\bar{a},F}^{\mathrm{anti-proton}} = -T_{a,F}^{\mathrm{proton}} \; .
\eeq
The asymmetries for $\pi^0$ production are between those
for $\pi^+$ and $\pi^-$. The same qualitative features persist
to RHIC energies, as can be seen from the comparison to the
STAR ($\pi^0$) and BRAHMS ($\pi^\pm$) data in Fig.~\ref{fig5b}.

For the second fit, we allow also sea- and anti-quark
$T_F$ functions. We then find the following parameters:
\begin{eqnarray}
{\mathrm{Fit\; II:}\;\;\;\;}
&&  N_{u_v} = 0.353  \; , \;\;\; N_{d_v} = -0.594   \; , \nonumber \\
&&N_{\bar{u}} = -N_{\bar{d}}=- N_{u_{\mathrm{sea}}}=
N_{d_{\mathrm{sea}}}=-19.8  \; , \;\;\; N_{s} =
N_{\bar{s}}=-6.63 \; , \nonumber \\
&& \alpha_{u_v} =  0.696 \; , \;\;\;  \beta_{u_v} = 0.559 \; , \;\;\;
\alpha_{d_v} = 0.312 \; , \;\;\; \beta_{d_v} = 0.488 \; , \nonumber \\
&& \alpha_{\bar{u}}=\alpha_{u_{\mathrm{sea}}}=\alpha_{\bar{d}}=
\alpha_{d_{\mathrm{sea}}}=\alpha_s=\alpha_{\bar{s}} = 2.91 \; , \nonumber \\
&&\beta_{\bar{u}}=\beta_{u_{\mathrm{sea}}}=\beta_{\bar{d}}=
\beta_{d_{\mathrm{sea}}}=\beta_s=\beta_{\bar{s}}  = 0.351 \; .
\label{fitres2}
\end{eqnarray}
Here the relations among the various parameters for sea and
anti-quarks are not fit results, but have been imposed. As before,
we have a normalization factor $N_{\mathrm{E704}}=0.5$ for the
calculated theory asymmetries at the E704 kinematics.
The results of this fit are also shown in Figs.~\ref{fig5a}
and~\ref{fig5b}, by the dashed lines. One can see that the fit
is rather similar to Fit~I, but does slightly better. Indeed, the
fit has $\chi^2=292.6$.
\begin{figure}[t]
%\hspace*{-1cm}
%\epsfig{figure=star-610p.eps,width=0.45\textwidth}
%\epsfig{figure=bramhs-610.eps,width=0.43\textwidth}
\epsfig{figure=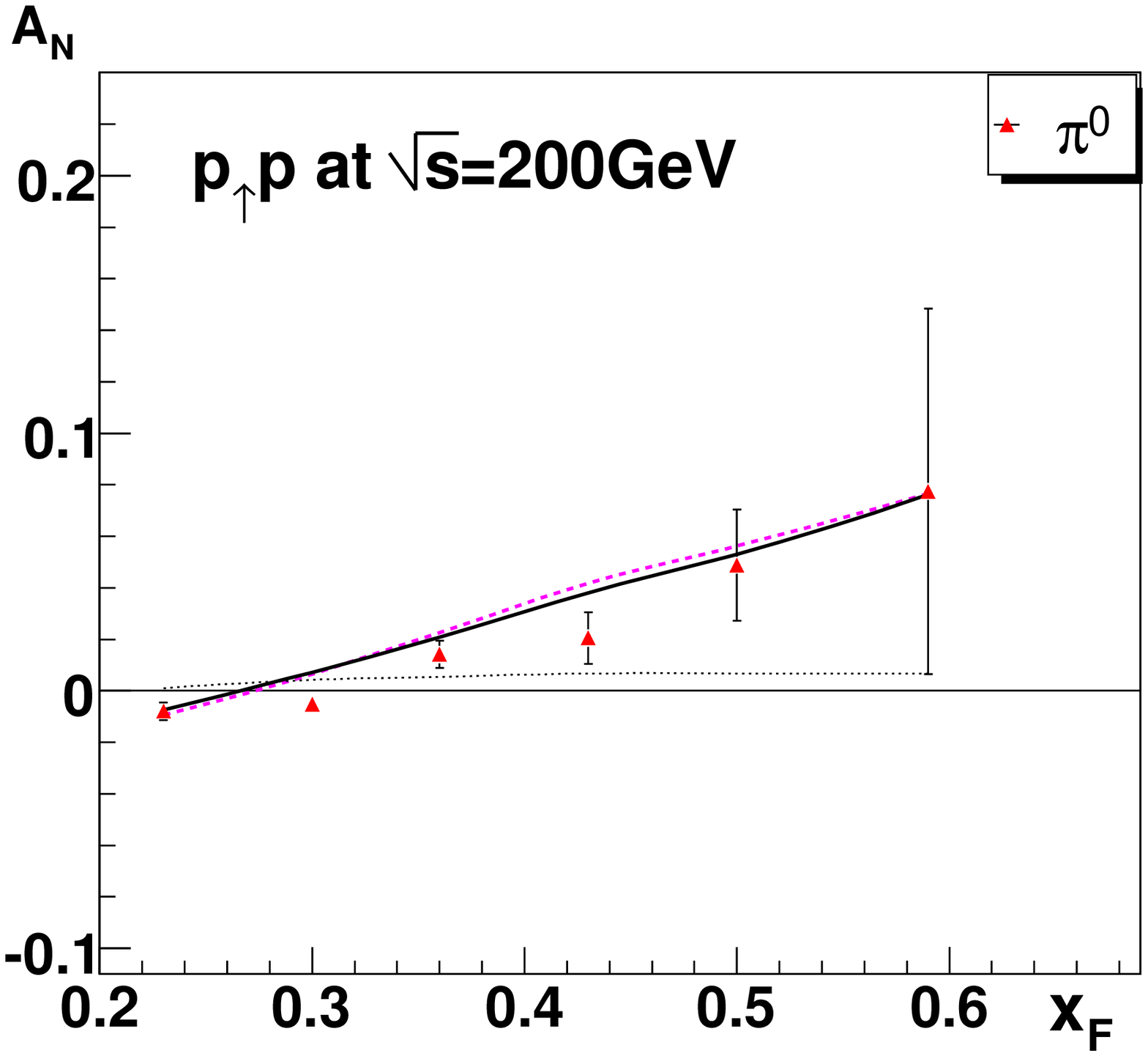,width=8cm, height=7cm}
\epsfig{figure=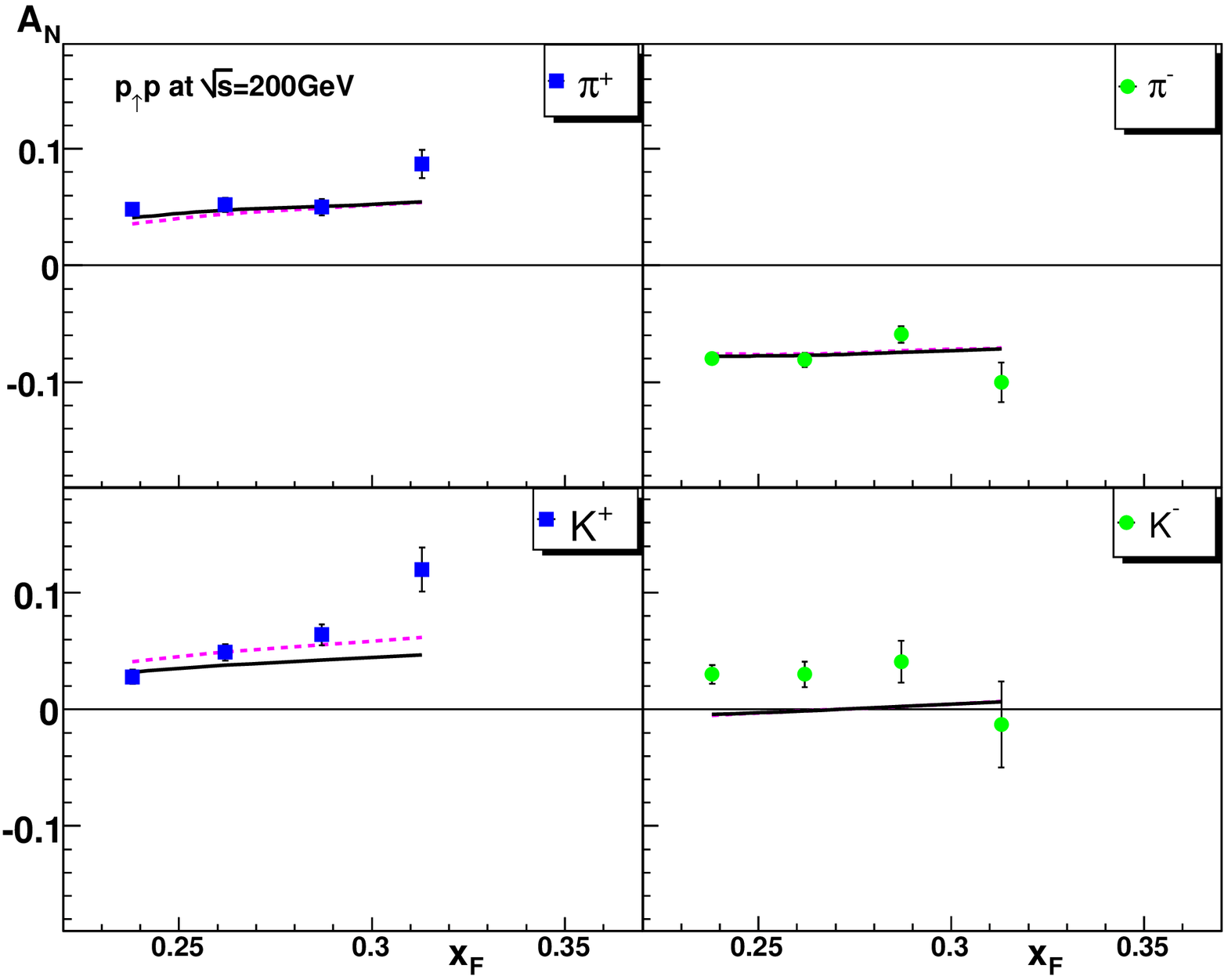,width=8cm, height=7cm}

\caption{\it Comparison of the single-spin asymmetries $A_N$ using
our fit results in Eqs.~(\ref{ansatz}),(\ref{fitres1}),(\ref{fitres2}) 
to the preliminary RHIC data by the STAR~\cite{star} (left) and
BRAHMS~\cite{brahms} (right) collaborations. The solid lines are
for Fit~I (Eq.~(\ref{fitres1})), and the dashed ones are for
Fit~II (Eq.~(\ref{fitres2})). The lower dotted line in the figure
shows the contribution to $A_N$ by the ``non-derivative'' terms
alone, for Fit~I.}  \label{fig5b}
\end{figure}
While the valence-quark densities completely dominate for the
fixed-target case, the sea distributions play a somewhat more
significant role at RHIC. We note, however, that in the case
of $A_N$ in $K^-$ production (see Fig.~\ref{fig5b}) even our
fit with a sea distribution does not lead to a significant change
in the theoretical result. This is surprising at first sight,
because the $K^-$ has no valence quarks in common with the proton,
so that sea quarks and anti-quarks should be particularly
important here. We found that the precise admixture of
valence (``favored''), non-valence (``un-favored''), and gluon
fragmentation functions is very relevant in this case, as well as
that of the hard-scattering functions. We could improve the
description of $A_N$ in $K^-$ production only by assuming a
very large negative correlation function $T_{\bar{u},F}$.

We also address the numerical relevance of the ``non-derivative''
terms that we have calculated in this work. The dotted lines
in the upper left part of Fig.~\ref{fig5a} and in the left part
of Fig.~\ref{fig5b} show the contributions to $A_N$ that one obtains
from the ``non-derivative'' terms alone, for the case of the two-flavor
fit (Fit~I). One can see that these contributions are of relatively
moderate ($\sim \mathrm{few}\,\%$) size, but non-negligible. They
play a bigger role at RHIC energies.
There is roughly a 25\% increase in the value of $\chi^2$ when
the ``non-derivative'' contributions are neglected.
Of course, one could refit the $T_{a,F}$ distributions without the
``non-derivative'' contributions, in which case the theoretical
spin asymmetry would be again very close to the dashed or solid lines
in the figures. However, we found that such a fit has a slightly
worse $\chi^2$, and in any case it leads to a fairly different
set of $T_{a,F}$ distributions.

\begin{figure}[t]
\hspace*{-1cm}
\epsfig{figure=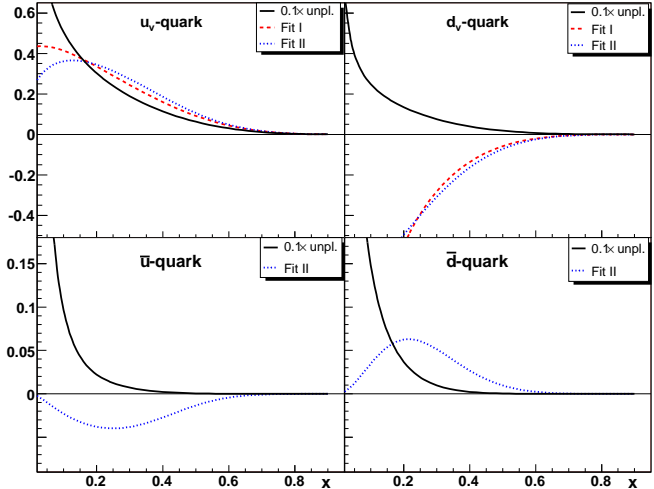,width=0.55\textwidth}
\caption{\it $T_{a,F}$ distributions for $a=u_v,d_v,\bar{u},\bar{d}$
resulting from our fits in Eqs.~(\ref{fitres1}) and~(\ref{fitres2}),
at scale $\mu=2$~GeV. We also show the corresponding unpolarized parton
distribution functions~\cite{cteq5l}, scaled by 1/10.}  \label{tfplot}
\end{figure}

In Fig.~\ref{tfplot} we show the $T_{a,F}$ distributions that
we have found in our fits, for $u$ and $d$ valence- and anti-quarks,
at scale $\mu=2$~GeV.
The dashed lines are for the ``two-flavor'' Fit~I, while the
dotted ones are for Fit~II. As one can see, the valence distributions
are rather similar in the two fits. Only Fit~II has anti-quark
distributions. For all distributions, we also show the corresponding
unpolarized leading-twist densities, scaled by 1/10 for better visibility.

It is interesting to speculate about the reasons why the overall
quality of our fits is relatively poor. We first remind the
reader that for the reasons discussed earlier we have rescaled
all theory asymmetries in the kinematic region of the E704 data
by a factor $1/2$ in the fit.  Without the rescaling factor,
the total $\chi^2$ of the fit would be increased by almost 100 units from
the current $\sim 300$, while the sign and the general shape of the 
asymmetries would still be consistent with the data.  Small changes
in the normalization of the RHIC data sets do not lead to a very
significant further reduction of $\chi^2$. We also found
that an even better description of all RHIC data is possible
if one excludes the E704 data from the fit. Such a fit then tends to
badly describe the $A_N$ data from E704, even when a normalization
factor is applied to the latter. We recall once more
that the E704 data are in a kinematic regime where the theoretical
calculation of even the unpolarized cross section is challenging, and
that we set $\ell_\perp=1.2$~GeV for them. It is therefore perhaps
not surprising that we find that the consistency of the total
data set for $A_N$ appears limited. We also caution the reader, however,
that much of the RHIC data is still preliminary and one needs to await
further experimental information before drawing final conclusions.

\begin{figure}[b]
\hspace*{-0.5cm}
\includegraphics[height=7.8cm,angle=90]{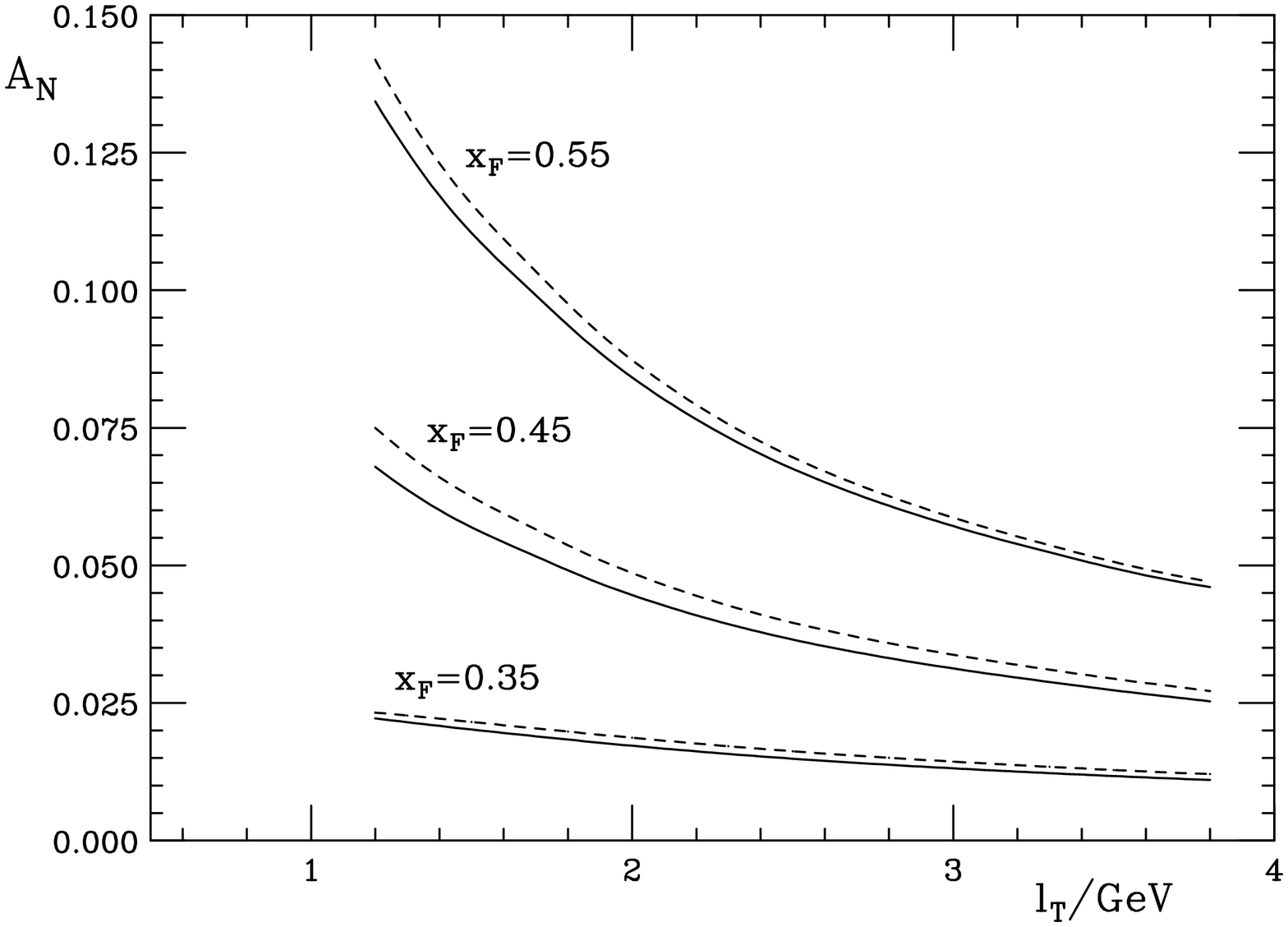}
\includegraphics[height=7.8cm,angle=90]{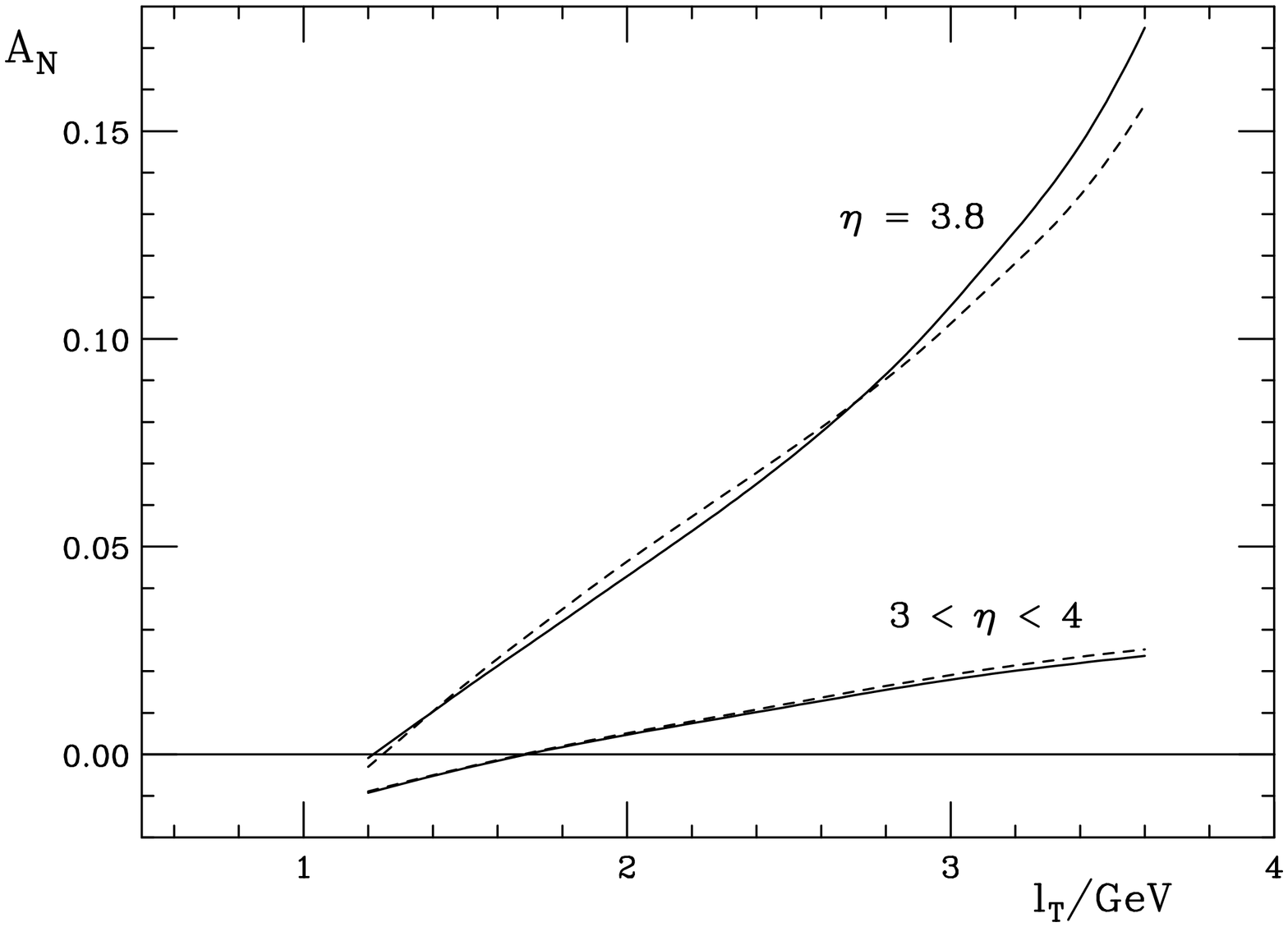}
\vspace*{-0.2cm}
\epsfysize=4.2in \caption{\it (a) Dependence of $A_N$ for
$\pi^0$ production at RHIC at $\sqrt{S}=200$~GeV
on $\ell_\perp$, for three different
values of $x_F$. Solid lines are for the $T_{a,F}$ distributions
of Fit~I, dashed ones are for Fit~II.
(b) Dependence on $\ell_\perp$ for fixed pion
pseudo-rapidity $\eta=3.8$ and when taking an average over
the bin $3< \eta<4$.}  \label{fig6}
\end{figure}
\begin{figure}[t]
\hspace*{-0.5cm}
\includegraphics[height=9cm,angle=90]{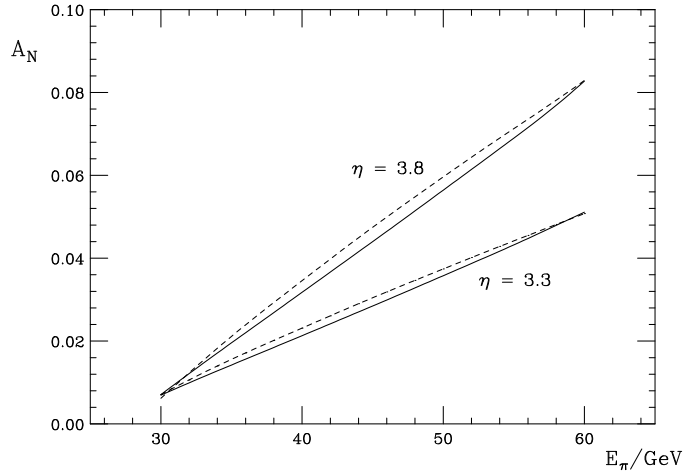}
\vspace*{-0.2cm}
\epsfysize=4.2in \caption{\it (a) Dependence of $A_N$ for
$\pi^0$ production at RHIC at $\sqrt{S}=200$~GeV
on the pion energy $E_\pi$,
for fixed $\eta=3.3$ and $\eta=3.8$. Solid lines are for the
$T_{a,F}$ distributions of Fit~I, dashed ones are for Fit~II.}
\label{fig6b}
\end{figure}

We now use our fitted twist-three correlation functions $T_{a,F}$
of Eqs.~(\ref{ansatz}),(\ref{fitres1}), and (\ref{fitres2})
to make a set of further predictions that may be tested at RHIC. The
first one concerns the dependence of $A_N$ on the produced hadron's
transverse momentum $\ell_\perp$. This is a particularly
interesting observable, given the power-suppressed nature
of $A_N$. In fact, as we discussed in the Introduction,
$A_N$ is expected to decrease as $1/\ell_\perp$, at a given
$x_F$. In Fig.~\ref{fig6}(a) we plot $A_N$ for $\pi^0$ production
at $\sqrt{S}=200$~GeV at three fixed values of the Feynman variable,
$x_F=0.35,\, 0.45,\, 0.55$, for our two sets of $T_{a,F}$ in
Fit~I and Fit~II. One can clearly see the fall-off
with $\ell_\perp$. In order to experimentally verify this fall-off,
that is, to keep $x_F$ fixed while varying $\ell_\perp$, one
would need to vary the scattering angle. On the other hand,
if measurements are made at a (roughly) fixed scattering angle $\theta$
or pseudo-rapidity $\eta=-\ln\tan(\theta/2)$, $x_F$ will increase along with
$\ell_\perp$, as seen from the relation
\beq
x_F = \frac{2 \ell_\perp}{\sqrt{S}} \sinh(\eta) \; .
\label{xfpt}
\eeq
This is often the experimentally more relevant situation.
As one can see from Fig.~\ref{fig6}(a), even though $A_N$
decreases with $\ell_\perp$ at fixed $x_F$, its {\it increase}
with increasing $x_F$ at a given $\ell_\perp$ appears to be stronger.
Hence, one expects that for measurements at fixed scattering angle
$A_N$ will actually increase with $\ell_\perp$. Indeed, this is
the case, as shown in Fig.~\ref{fig6}(b). If one averages experimentally
over a bin of forward rapidities, say, $3< \eta<4$, the increase
of $A_N$ with $\ell_\perp$ is less pronounced but still there.
Another observable, most relevant to measurements at STAR~\cite{cross_star},
is the dependence on the asymmetry on the pion energy
$E_\pi=\ell_\perp\, \cosh(\eta)$ for fixed $\eta$. This is plotted
in Fig.~\ref{fig6b} for our two fits, for the cases $\eta=3.3$ and
$\eta=3.8$.

\begin{figure}[t]
\hspace*{-0.5cm}
\includegraphics[height=9cm,angle=90]{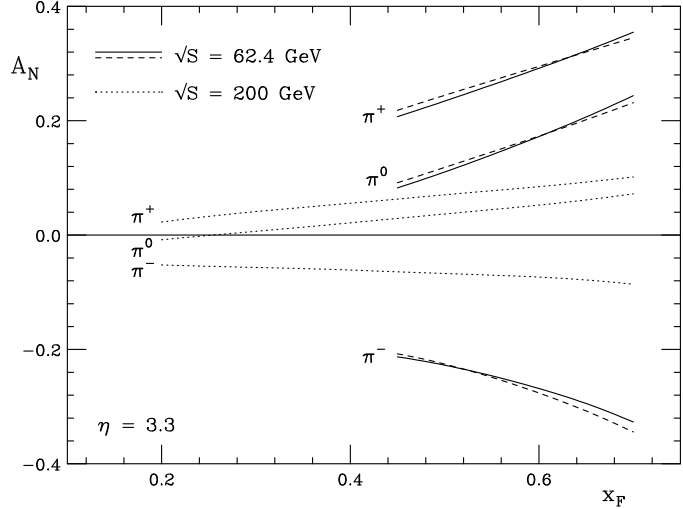}
\vspace*{-0.2cm}
\epsfysize=4.2in \caption{\it Comparison of $A_N$ for $\pi^\pm$ and
$\pi^0$ production as functions of $x_F$ at $\sqrt{S}=62.4$ and at 200~GeV,
using our two fit results. We have chosen a fixed pion pseudo-rapidity
$\eta=3.3$.}
\label{fig7}
\end{figure}

It is also interesting to consider the energy dependence
of $A_N$. So far, we have data from fixed-target scattering
at $\sqrt{S}\sim 20$~GeV and from RHIC at a center-of-mass
energy about an order of magnitude higher. In order to shed
further light on the mechanisms responsible for the large observed
values of $A_N$ at these two energies, information at an intermediate
energy will be particularly useful. In the 2006 run, data have been
taken at RHIC at $\sqrt{S}=62.4$~GeV. Using our above fit results,
we find the theoretical expectations for $A_N$ for $\pi^\pm$ and
$\pi^0$ production
as functions of $x_F$ shown in Fig.~\ref{fig7}. We have for now again
correlated $x_F$ and $\ell_\perp$ through Eq.~(\ref{xfpt}) at
fixed pseudo-rapidity $\eta=3.3$. Clearly, comparison
to eventual data will require implementation of the correct
kinematics. In the figure, we compare results for
$\sqrt{S}=62.4$~GeV and 200~GeV. One can see that, at fixed
$x_F$ and $\eta$, a significant increase of $A_N$ with
energy should be expected.

\begin{figure}[t]
\hspace*{-0.5cm}
\includegraphics[height=9cm,angle=90]{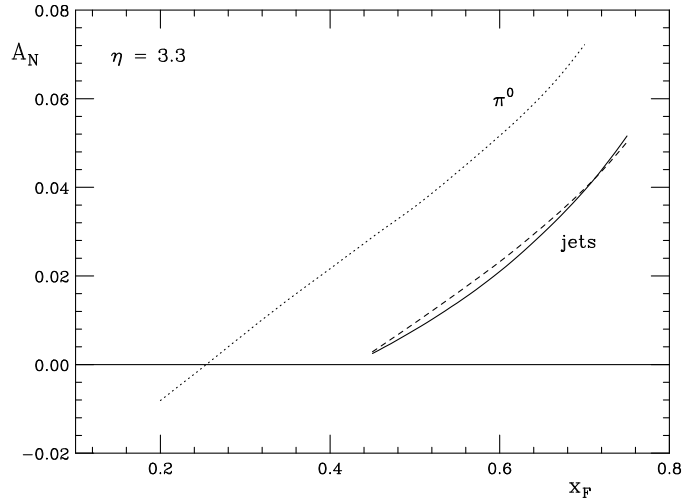}
\vspace*{-0.2cm}
\epsfysize=4.2in \caption{\it Single-spin asymmetry for
jet production (with transverse momentum $\ell_\perp>3$~GeV)
at RHIC at $\sqrt{S}=200$~GeV, as a
function of $x_F$ for fixed pseudo-rapidity $\eta=3.3$. We
show the results for both Fit~I (solid) and Fit~II (dashed).
The dotted curve shows the same for $\pi^0$ production for Fit~I.}
\label{fig8a}
\end{figure}

We finally briefly turn to processes other than inclusive-hadron
production. We first consider the spin asymmetry in single-inclusive
jet production. The partonic hard-scattering functions for this
case are the same as for hadron production, but there are no
fragmentation functions here. The result for $A_N$ for jet-production
at RHIC at forward $x_F$ and fixed $\eta=3.3$ is shown
in Fig.~\ref{fig8a}. For comparison, we also show again the
corresponding curve for $\pi^0$ production for Fit~I. One
can see that essentially the asymmetry for jets is shifted by
a factor $\sim 2$ to the right with respect to that for pions.
This can be understood from the fact that in the kinematic
regime relevant here a pion takes on average roughly 50\% of a
fragmenting parton's energy~\cite{strik}, whereas {\it all} of
the energy goes into a jet.

Another process of interest is prompt-photon production.
Photons are generally much less copiously produced at
RHIC energies than pions, which results in larger statistical
uncertainties on the spin asymmetries. However, given the
progress on luminosity and polarization at RHIC, first measurements
of $A_N$ for prompt photons should become possible in the near-term
future. Photons have the advantage that one important production
mechanism is quark-gluon Compton scattering, $qg\to \gamma q$,
with the reaction  $q\bar{q}\to \gamma g$ yielding a smaller
contribution. In addition, in these processes the photon couples
in a ``direct'' (or, point-like) way, that is, there are no fragmentation
functions involved. Photons can, however, also be produced in jet
fragmentation~\cite{owens}. The relative importance of the ``direct''
and the fragmentation contributions depends on kinematics, but also
on aspects of the experimental measurement. It is possible, for example,
to largely suppress the fragmentation contribution by a so-called
photon isolation cut~\cite{owens}. In the following, in order to obtain first
estimates, we will calculate the single-spin asymmetry for prompt
photons based on either the ``direct'' contributions alone, or
on the sum of the ``direct'' and the full fragmentation contributions.
The former is more representative of the asymmetry for an isolated
photon cross section, while the latter corresponds to a fully inclusive
measurement. When data will become available, a more careful theoretical
analysis will clearly become necessary.

Predictions for $A_N$ for prompt-photon production can then
be obtained from Eq.~(\ref{finalcr})
by using the appropriate hard-scattering functions for
the reactions $(qg)g\to \gamma q$ and $(qg)\bar{q}\to \gamma g$
and by replacing the coupling factor $\alpha_s^2$ by
$\alpha_s\, \alpha_{\mathrm{e.m.}}\, e_q^2$,
where $\alpha_{\mathrm{e.m.}}$
is the electromagnetic coupling constant
and $e_q$ is the fractional electric charge carried by the quark
of flavor $q$.
We give the resulting expressions in Appendix~B, along with
the corresponding ones for the unpolarized case, to be used
in Eq.~(\ref{finalcrU}), with the same replacement of the couplings.
Taking the fit results of
\begin{figure}[b]
\hspace*{-0.5cm}
\includegraphics[height=9cm,angle=90]{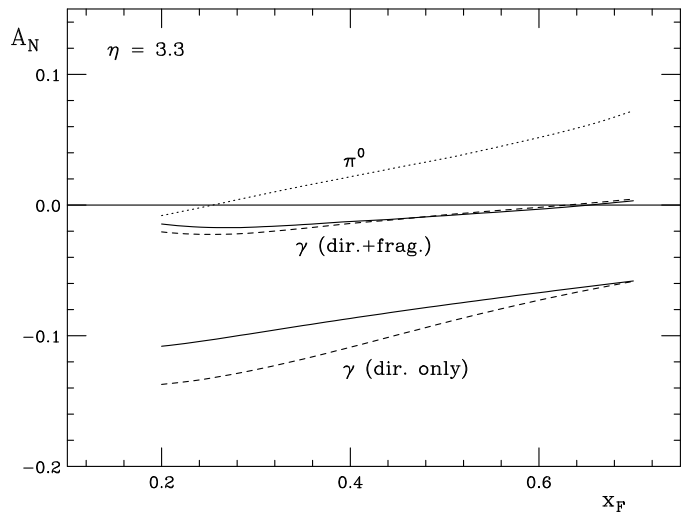}
\vspace*{-0.2cm}
\epsfysize=4.2in \caption{\it Single-spin asymmetry for
prompt-photon production at RHIC at $\sqrt{S}=200$~GeV, as a
function of $x_F$ for fixed pseudo-rapidity $\eta=3.3$. We
show the predictions for both Fit~I (solid) and Fit~II (dashed).
We show separately the results for the cases when the fragmentation
component is taken into account or neglected. The dotted curve shows the
earlier result for $\pi^0$ production for Fit~I.}
\label{fig8}
\end{figure}
Eqs.~(\ref{ansatz}),(\ref{fitres1}),(\ref{fitres2}) we obtain the predictions
shown in Fig.~\ref{fig8}, which are for $\sqrt{S}=200$~GeV.
Again we have chosen a fixed pseudo-rapidity $\eta=3.3$.
For comparison we also show again the corresponding
result for $A_N$ for $\pi^0$ production. The differences
between the asymmetries for photons (``direct only'') and
$\pi^0$ are quite striking. They mostly result from a rather different
structure of the corresponding hard-scattering functions
(see Eqs.~(\ref{qgg}) and~(\ref{qgga}) in Appendices A and B,
respectively) and are therefore a real prediction of the formalism.
One also sees that the two fits I and II give somewhat different
predictions for $A_N$ for the photons ``direct only'' case.
This is due to the contributions from $(\bar{q}g)q$ scattering which
are present for Fit II, but absent for fit~I since for this fit
we assumed that the sea quark $T_{a,F}$ functions vanish.

When the fragmentation contribution to the prompt photon
cross section is taken into account, the single-spin asymmetry
becomes much more like the one for $\pi^0$, at least for the lower
$x_F$. The reason is that in the kinematic regime relevant here,
i.e. at relatively low transverse momenta $\ell_\perp$, the fragmentation
component actually dominates the cross section. Future measurements
of the single-spin asymmetry for isolated photons should see
an asymmetry close to the lower one (``direct only'') in Fig.~\ref{fig8}, 
while for the fully inclusive (non-isolated) case $A_N$ should be
smaller and closer to that for $\pi^0$ production.

\section{Conclusions and outlook \label{concl}}

We have presented a new study of the single-spin asymmetry
in single-inclusive hadron production in hadronic scattering.
The importance of this asymmetry lies in the new insights into
nucleon structure it may provide, but also in the challenge that
its description poses for QCD theory due its power-suppressed
nature. We have extend the previous calculations in~\cite{qs} by
deriving also the so-called ``non-derivative'' contributions to
the spin-dependent cross section. We have found that these combine
with the ``derivative'' pieces into a remarkably simple structure.

Using our derived cross section, we have also made first
phenomenological studies, using the E704 fixed-target and
the latest preliminary RHIC (STAR and BRAHMS) data. We have
found that a simultaneous description of all these data
is possible, albeit at a more qualitative, than quantitative,
level, with the RHIC data overall better described. The
``non-derivative'' contributions we have calculated are of
moderate importance. 

We have finally made predictions for a number
of other single-spin observables at RHIC, in particular for
the $\ell_\perp$-dependence of $A_N$ for $\pi^0$ production, 
for scattering at 62.4~GeV, 
and for the asymmetries for single-jet and prompt-photon 
final states.

For the future, it will be desirable to extend our work in a number
of ways. Regarding the theoretical framework, one should eventually
include also purely gluonic higher-twist correlation functions. These
are expected to be of particular relevance for the spin asymmetry
at mid-rapidity, which was found experimentally to be small~\cite{phenix}.
Also, we have so far only considered the ``soft-gluon'' contributions
to the spin asymmetry, for which the gluon in the twist-three
quark-gluon correlation function is soft. As we mentioned earlier,
there are also in general ``soft-fermion'' contributions. These
involve, among other things, the functions $T_{a,F}(x,0)$, rather
than the $T_{a,F}(x,x)$ that we found for the soft-gluon case.
The soft-fermion contributions have their own hard-scattering
functions and may make a significant contribution to the spin
asymmetry as well. Further points of interest will be the
evolution of the functions $T_{a,F}(x,x)$, and the detailed
study of similarities and differences between our approach and
the formalism of~\cite{ams}, where the spin asymmetry in the process
$p^{\uparrow}p\to \pi \pi X$ has been considered in the context of
gauge links in hard-scattering processes. This could be achieved for
example by a study of the single-spin asymmetry for two-pion or
two-jet production in the framework of~\cite{qs} that we have used here.

Regarding phenomenology, our studies so far have a more
illustrative character. With new experimental information
arriving from RHIC, however, we will be entering an era
where detailed global analyses of the data on $A_N$ will
become possible.

\section*{Acknowledgments}
We are grateful to L.\ Bland, D.\ Boer, C.\ Bomhof, Y.\ Koike, P.\ Mulders,
and G.\ Sterman for useful discussions. C.K. is supported by the Marie Curie
Excellence Grant under contract MEXT-CT-2004-013510.
J.Q. is supported in part by the U. S.
Department of Energy under grant No. DE-FG02-87ER-40371. W.V. and
F.Y. are finally grateful to RIKEN, Brookhaven National
Laboratory and the U.S. Department of Energy (contract number
DE-AC02-98CH10886) for providing the facilities essential for the
completion of their work.

\section*{Appendix A: Hard-scattering functions for inclusive-hadron
production}
In this Appendix we list the hard-scattering
functions relevant for single-inclusive hadron production. For
each partonic channel, we give the functions $H^I_{ab\rightarrow c}$
and $H^F_{ab\rightarrow c}$, which are to be used in Eq.~(\ref{Hfin}).
We also present the corresponding unpolarized cross sections 
$H^U_{ab\rightarrow c}$ for Eq.~(\ref{finalcrU}). We have:

\noindent
$qg\to qg$ scattering:
\beeq
H^U_{qg\rightarrow q}(\hat{s},\hat{t},\hat{u})&=&
\frac{C_F}{N_C} \left[ -\frac{\hat{s}}{\hat{u}}-
\frac{\hat{u}}{\hat{s}}\right] \;
\left[ 1 - \frac{N_C}{C_F} \frac{\hat{s}\hat{u}}{\hat{t}^2} \right]
\; , \nonumber \\
H^I_{qg\rightarrow q}(\hat{s},\hat{t},\hat{u})&=&
\frac{1}{2(N_C^2-1)} \left[ -\frac{\hat{s}}{\hat{u}}-
\frac{\hat{u}}{\hat{s}}\right] \;
\left[ 1 - N_C^2 \frac{\hat{u}^2}{\hat{t}^2} \right] \; , \nonumber \\
H^F_{qg\rightarrow q}(\hat{s},\hat{t},\hat{u})&=&
\frac{1}{2N_C^2 (N_C^2-1)} \left[ -\frac{\hat{s}}{\hat{u}}-
\frac{\hat{u}}{\hat{s}}\right] \;
\left[ 1 + 2 N_C^2 \frac{\hat{s}\hat{u}}{\hat{t}^2} \right]  \; .
\eeeq

\noindent
$qg\to gq$ scattering:
\beeq
H^U_{qg\rightarrow g}(\hat{s},\hat{t},\hat{u})&=&
\frac{C_F}{N_C}  \left[ -\frac{\hat{s}}{\hat{t}}-
\frac{\hat{t}}{\hat{s}}\right] \;
\left[ 1 - \frac{N_C}{C_F}  \frac{\hat{s}\hat{t}}{\hat{u}^2} \right]
\; , \nonumber \\
H^I_{qg\rightarrow g}(\hat{s},\hat{t},\hat{u})&=&
\frac{1}{2(N_C^2-1)} \left[ -\frac{\hat{s}}{\hat{t}}-
\frac{\hat{t}}{\hat{s}}\right] \;
\left[ 1 - N_C^2 \frac{\hat{t}^2}{\hat{u}^2} \right]\; ,\nonumber \\
H^F_{qg\rightarrow g}(\hat{s},\hat{t},\hat{u})&=&
-\frac{1}{2(N_C^2-1)} \left[ -\frac{\hat{s}}{\hat{t}}-
\frac{\hat{t}}{\hat{s}}\right] \;
\left[ 1 - N_C^2 \frac{\hat{s}^2}{\hat{u}^2} \right] \; .
\label{qgg}
\eeeq

\noindent
$q\bar{q}\to gg$ scattering:
\beeq
H^U_{q\bar{q}\rightarrow g}(\hat{s},\hat{t},\hat{u})&=&
\frac{2 C_F^2}{N_C} \left[ \frac{\hat{t}}{\hat{u}} +
\frac{\hat{u}}{\hat{t}} \right]
\left[ 1- \frac{N_C}{C_F} \frac{\hat{t}\hat{u}}{\hat{s}^2}
\right] \; , \nonumber \\
H^I_{q\bar{q}\rightarrow g}(\hat{s},\hat{t},\hat{u})&=&
-\frac{1}{2 N_C^3}\left[ \frac{\hat{t}}{\hat{u}} +
\frac{\hat{u}}{\hat{t}} \right]
\left[ 1+2 N_C^2  \frac{\hat{t}\hat{u}}{\hat{s}^2}\right]\; ,\nonumber \\
H^F_{q\bar{q}\rightarrow g}(\hat{s},\hat{t},\hat{u})&=&
-\frac{1}{2N_C}\left[ \frac{\hat{t}}{\hat{u}} +
\frac{\hat{u}}{\hat{t}} \right]
\left[ 1 - N_C^2 \frac{\hat{u}^2}{\hat{s}^2}\right]\; .
\eeeq

\noindent
$qq'\to qq'$ scattering:
\beeq
H^U_{qq'\rightarrow q}(\hat{s},\hat{t},\hat{u})&=&
\frac{C_F}{N_C} \left[ \frac{\hat{s}^2+\hat{u}^2}{\hat{t}^2} \right]
\; , \nonumber \\
H^I_{qq'\rightarrow q}(\hat{s},\hat{t},\hat{u})&=&
-\frac{1}{N_C^2}\left[ \frac{\hat{s}^2+\hat{u}^2}{\hat{t}^2} \right]\; ,
\nonumber \\
H^F_{qq'\rightarrow q}(\hat{s},\hat{t},\hat{u})&=&
-\frac{1}{2N_C^2}\left[ \frac{\hat{s}^2+\hat{u}^2}{\hat{t}^2} \right]\; .
\eeeq

\noindent
$qq'\to q'q$ scattering:
\beeq
H^U_{qq'\rightarrow q'}(\hat{s},\hat{t},\hat{u})&=&
\frac{C_F}{N_C} \left[ \frac{\hat{s}^2+\hat{t}^2}{\hat{u}^2} \right]
\; , \nonumber \\
H^I_{qq'\rightarrow q'}(\hat{s},\hat{t},\hat{u})&=&
-\frac{1}{N_C^2}\left[ \frac{\hat{s}^2+\hat{t}^2}{\hat{u}^2} \right]\; ,
\nonumber \\
H^F_{qq'\rightarrow q'}(\hat{s},\hat{t},\hat{u})&=&
\frac{N_C^2-2}{2N_C^2}\left[ \frac{\hat{s}^2+\hat{t}^2}{\hat{u}^2} \right]\; .
\eeeq

\noindent
$qq\to qq$ scattering:
\beeq
H^U_{qq\rightarrow q}(\hat{s},\hat{t},\hat{u})&=&
\frac{C_F}{N_C}\left[ \frac{\hat{s}^2+\hat{u}^2}{\hat{t}^2}
+\frac{\hat{s}^2+\hat{t}^2}{\hat{u}^2} - \frac{2}{N_C}
\frac{\hat{s}^2}{\hat{t}\hat{u}}\right] \; , \nonumber \\
H^I_{qq\rightarrow q}(\hat{s},\hat{t},\hat{u})&=&
-\frac{1}{N_C^2}\left[ \frac{\hat{s}^2+\hat{u}^2}{\hat{t}^2} +
 \frac{\hat{s}^2+\hat{t}^2}{\hat{u}^2} - \frac{N_C^2+1}{N_C}
\frac{\hat{s}^2}{\hat{t}\hat{u}}\right]\; ,
\nonumber \\
H^F_{qq\rightarrow q}(\hat{s},\hat{t},\hat{u})&=&
-\frac{1}{2N_C^2}\left[ \frac{\hat{s}^2+\hat{u}^2}{\hat{t}^2} \right]+
\frac{N_C^2-2}{2N_C^2}\left[ \frac{\hat{s}^2+\hat{t}^2}{\hat{u}^2} \right]+
\frac{1}{N_C^3}\frac{\hat{s}^2}{\hat{t}\hat{u}}\; .
\eeeq

\noindent
$q\bar{q}\to q'\bar{q}'$ scattering:
\beeq
H^U_{q\bar{q}\rightarrow q'}(\hat{s},\hat{t},\hat{u})&=&
\frac{C_F}{N_C}\left[ \frac{\hat{t}^2+\hat{u}^2}{\hat{s}^2}
\right]\; , \nonumber \\
H^I_{q\bar{q}\rightarrow q'}(\hat{s},\hat{t},\hat{u})&=&
\frac{1}{2N_C^2}\left[ \frac{\hat{t}^2+\hat{u}^2}{\hat{s}^2} \right]\; ,
\nonumber \\
H^F_{q\bar{q}\rightarrow q'}(\hat{s},\hat{t},\hat{u})&=&
\frac{N_C^2-2}{2N_C^2}\left[ \frac{\hat{t}^2+\hat{u}^2}{\hat{s}^2} \right]\; .
\eeeq

\noindent
$q\bar{q}\to q\bar{q}$ scattering:
\beeq
H^U_{q\bar{q}\rightarrow q}(\hat{s},\hat{t},\hat{u})&=&
\frac{C_F}{N_C}\left[ \frac{\hat{t}^2+\hat{u}^2}{\hat{s}^2} +
\frac{\hat{s}^2+\hat{u}^2}{\hat{t}^2} - \frac{2}{N_C}
\frac{\hat{u}^2}{\hat{s}\hat{t}}\right]
\; ,\nonumber \\
H^I_{q\bar{q}\rightarrow q}(\hat{s},\hat{t},\hat{u})&=&
-\frac{N_C^2-2}{2N_C^2}\frac{\hat{s}^2+\hat{u}^2}{\hat{t}^2} +
\frac{1}{2N_C^2}\frac{\hat{t}^2+\hat{u}^2}{\hat{s}^2} -
\frac{1}{N_C^3} \frac{\hat{u}^2}{\hat{s}\hat{t}}\; ,
\nonumber \\
H^F_{q\bar{q}\rightarrow q}(\hat{s},\hat{t},\hat{u})&=&
-\frac{1}{2N_C^2}\left[ \frac{\hat{s}^2+\hat{u}^2}{\hat{t}^2} \right]+
\frac{N_C^2-2}{2N_C^2}\left[ \frac{\hat{t}^2+\hat{u}^2}{\hat{s}^2} \right]+
\frac{1}{N_C^3} \frac{\hat{u}^2}{\hat{s}\hat{t}}\; ,
\eeeq
where $C_F=(N_C^2-1)/2N_C=4/3$.

\section*{Appendix B: Hard-scattering functions for direct-photon production}
In this Appendix we list the hard-scattering
functions relevant for single-inclusive prompt-photon production. In this
case, there are only initial-state contributions $H^I_{ab\rightarrow \gamma}$
in Eq.~(\ref{Hfin}). For convenience, we also again give the unpolarized
contributions $H^U_{ab\rightarrow \gamma}$. We have:

\noindent
$qg\to \gamma q$ scattering:
\beeq
H^U_{qg\rightarrow \gamma}(\hat{s},\hat{t},\hat{u})&=&
\frac{e_q^2}{N_C}\left[ -\frac{\hat{s}}{\hat{t}}-
\frac{\hat{t}}{\hat{s}}\right] \;  , \nonumber \\
H^I_{qg\rightarrow \gamma}(\hat{s},\hat{t},\hat{u})&=&e_q^2
\frac{N_C}{N_C^2-1} \left[ -\frac{\hat{s}}{\hat{t}}-
\frac{\hat{t}}{\hat{s}}\right] \;  .
\label{qgga}
\eeeq

\noindent
$q\bar{q}\to \gamma g$ scattering:
\beeq
H^U_{q\bar{q}\rightarrow \gamma}(\hat{s},\hat{t},\hat{u})&=&
e_q^2 \frac{2 C_F}{N_C}\left[ \frac{\hat{t}}{\hat{u}} +
\frac{\hat{u}}{\hat{t}} \right] \;  , \nonumber \\
H^I_{q\bar{q}\rightarrow \gamma}(\hat{s},\hat{t},\hat{u})&=&
-e_q^2 \frac{1}{N_C^2}\left[ \frac{\hat{t}}{\hat{u}} +
\frac{\hat{u}}{\hat{t}} \right] \; .
\eeeq

\end{document}